\documentclass[%
aps,
pre,
superscriptaddress,
tightenlines,
showpacs,showkeys,
a4paper,
12pt,
longbibliography,
notitlepage
]{revtex4-1}
\usepackage[english]{babel}
\usepackage{amssymb,amsmath,stmaryrd,array}

\usepackage{graphicx}

\usepackage{subfig}

\makeatletter

\newcommand{\avr}[1]{\ensuremath{\langle{#1}\rangle}}

\newcommand{\cnj}[1]{{#1}^{\ast}}

\newcommand{\pdrs}[1]{\partial_{#1}}
\newcommand{\pdr}[2]{\frac{\partial #1}{\partial #2}}

\newcommand{\vdr}[2]{\dfrac{\delta #1}{\delta #2}}

\newcommand{\bnbl}{\boldsymbol{\nabla}}

\newcommand{\Tr}{\mathop{\rm Tr}\nolimits}

 \newcommand{\bs}[1]{\boldsymbol{#1}}
 \newcommand{\vc}[1]{\mathbf{#1}}
 \newcommand{\mvc}[1]{\mathbf{#1}}
 \newcommand{\uvc}[1]{\hat{\mathbf{#1}}}
 
 \newcommand{\ind}[1]{\mathrm{#1}}

\newcommand{\dd}{\mathrm{d}}

\newcommand{\st}{\mathrm{st}}

\newcommand{\trans}{\textit{trans}}
\newcommand{\cis}{\textit{cis}}

\makeatother

\begin{document}
\DeclareGraphicsExtensions{.eps,.png,.pdf}
\title{%
Photoinduced reordering in
thin azo-dye films and
light-induced reorientation dynamics of nematic 
liquid-crystal easy axis
}

\author{Alexei~D.~Kiselev}
\email[Email address: ]{kiselev@iop.kiev.ua}
\affiliation{%
 Institute of Physics of National Academy of Sciences of Ukraine, 
prospekt Nauki 46, 03028 Kiev, Ukraine}
\affiliation{%
Hong Kong University of Science and Technology, 
Clear Water Bay, Kowloon, Hong Kong}

\author{V.G.~Chigrinov}
\email[Email address: ]{eechigr@ust.hk}
\affiliation{%
Hong Kong University of Science and Technology,
Clear Water Bay, Kowloon,
Hong Kong}

\author{S.V.~Pasechnik}
\affiliation{%
Moscow State University of Instrument Engineering and Computer Science,
 Stromynka 20, 107846 Moscow, Russia}
\affiliation{%
Hong Kong University of Science and Technology, 
Clear Water Bay, Kowloon, Hong Kong}

\author{A.V.~Dubtsov}
\affiliation{%
Moscow State University of Instrument Engineering and Computer Science,
 Stromynka 20, 107846 Moscow, Russia}


\date{\today}

\begin{abstract}
We theoretically study the kinetics of photoinduced reordering 
triggered by linearly polarized (LP) reorienting light 
in thin azo-dye films that were initially illuminated with  
LP ultraviolet (UV) pumping beam. 
The  process of reordering is
treated as a rotational diffusion of molecules in 
the light intensity-dependent mean-field potential.
The two dimensional (2D) diffusion model 
which is based on the free energy rotational Fokker-Planck equation 
and describes the regime of in-plane
reorientation  is  generalized to analyze 
the dynamics of the azo-dye order parameter tensor
at varying polarization azimuth of the reorienting light.
It is found  that, in the photosteady state,
the intensity of LP reorienting light
determines
the scalar order parameter (the largest eigenvalue of the order
parameter tensor), whereas
the steady state orientation of
the corresponding eigenvector
(the in-plane principal axis)
depends solely on the polarization azimuth.
We show that, under certain conditions,
reorientation takes place only if 
the reorienting light intensity exceeds its critical value.
Such threshold behavior is predicted to occur in the bistability region
provided that the initial principal axis
lies in the polarization plane of reorienting light.
The model is used to interpret the experimental data on 
the light-induced azimuthal gliding of liquid-crystal easy axis on 
photoaligned azo-dye substrates.

\end{abstract}

 \maketitle

\section{Introduction}
\label{sec:intro}

Linearly polarized (LP) ultraviolet (UV) or visible light
is known to have a profound effect on 
the physical properties of some photosensitive materials
such as compounds containing azobenzene and its derivatives~\cite{Zhao:bk:2009}.
In particular, such materials
may become dichroic and
birefringent under the action of LPUV light.
Over the past few decades 
this phenomenon~---~the so-called 
effect
of  \textit{photoinduced optical anisotropy} 
(POA)~---~which has a long history dating 
back almost a century
to the original paper by Weigert~\cite{Wieg:1919}  
has been attracted much attention
for both technological and fundamental reasons.
POA is of primary importance in  
the development of tools dealing with the light-controlled anisotropy
and the materials that
exhibit POA are very promising for use in many
photonic
applications~\cite{Eich:1987,Nat:1992,Pras:1995,Blinov:chemph:1999}.

POA also lies at the heart of the \textit{photoalignment} (PA) technique 
which is employed in the manufacturing process of liquid crystal displays 
for fabricating high quality aligning substrates~\cite{Yang:bk:2006,Chigrin:bk:2008}.
This method  
avoids the drawbacks of the traditional mechanical surface treatment
and uses LPUV light to induce anisotropy of the angular
distribution of molecules in a photosensitive film~\cite{Kelly:jpd:2000,Chigrin:bk:2008}.

When the irradiated layer is
brought in contact with a liquid crystal (LC),
the surface ordering originated 
from the photoinduced anisotropy
manifests itself in
the anisotropic part of the surface 
tension known as the anchoring 
energy and determines
the anchoring characteristics such as 
the anchoring strengths and the \textit{easy axis},
$\vc{n}_e$, that defines the direction of
preferential orientation of LC molecules at the surface. 
In this way, the light can be used as a 
means to control the anchoring properties of 
photosensitive materials.

This is the light-induced anisotropy of the orientational distribution
of molecules that reveals itself in POA and can be described as 
the \textit{photoinduced orientational ordering}  in photosensitive materials. 
Such ordering, though not being understood very well,
can generally occur by a variety of photochemically induced processes.  
These typically may involve such transformations as photoisomerization, crosslinking,
photodimerization and photodecomposition 
(a recent review can be found in Refs.~\cite{Chigr:rewiev:2003,Chigrin:bk:2008}).

The mechanism of photoinduced ordering
underlying POA and related photoaligning properties therefore cannot be universal.
Rather they crucially depend on the material in question
and on a number of additional factors such as
irradiation conditions, surface interactions etc. 
These factors combined with the action of light
may result in 
different regimes of the photoinduced ordering kinetics
leading to the formation of various photoinduced orientational
structures.  

The photoalignment has been studied in a number of
polymer systems including dye doped polymer layers~\cite{Gibbon:nat:1991,Furum:1999},
cinnamate polymer
derivatives~\cite{Chig:jjap:1992,Dyad:jetpl:1992,Gal:1996,Barn:2000}
and side chain
polymers containing chemically linked
azochromophores 
(azopolymers)~\cite{Petry:1993,Holme:1996,Blin:1998,Iked:2000}.
POA observed in similar 
polymers~\cite{Eich:1987,Nat:1992,Holme:1996,Petry:1993,Wies:1992,Blin:1998,Kis:epj:2001}  
was found to be long term stable 
so that the photoinduced anisotropy does not disappear after switching off 
the irradiation.

Most of the above mentioned polymer systems represent
azocompounds that exhibit POA 
driven by the \trans-\cis\ photoisomerization.
In theoretical  
models~\cite{Ped:1997,Ped:1998,Puch:1998,Hvil:2001,Kis:epj:2001,Kis:jpcm:2002,Sekkat:jpcb:2002}, 
they are treated as ensembles of 
the stable \trans\ isomers characterized by
elongated rod-like molecular conformation and 
the bent banana-like shaped 
\cis\ isomers. 

The mechanism of photoisomerization implies that 
the key processes behind the photoinduced orientational ordering
of azo-dye molecules 
are photochemically induced \trans-\cis\  isomerization and subsequent
thermal and/or photochemical \cis-\trans\  back isomerization of
azobenzene chromophores. 
Owing to pronounced absorption dichroism of photoactive groups, 
the photoisomerization rate strongly
depends on orientation of the azo-dye molecules relative
to the polarization vector of the pumping light, $\vc{E}_{UV}$. 

When the \cis\ isomers are short-living, 
the \cis\ state becomes
temporary populated during photoisomerization but reacts immediately
back to the stable \trans\ isomeric form.
The
\trans-\cis-\trans\ isomerization cycles are accompanied by 
rotations of the azo-dye molecules that tend to 
minimize the absorption and 
become oriented along directions normal to 
the polarization vector of the exciting light
$\vc{E}_{UV}$
(the molecules with the optical transition dipole moment 
oriented perpendicular to $\vc{E}_{UV}$ are almost inactive). 
Non-photoactive groups may then undergo reorientation due to
cooperative 
motion~\cite{Holme:1996,Nat:1998,Puch:1998,Kis:jpcm:2002,Sekkat:jpcb:2002}.

From the above it might be concluded that
the photoinduced orientational structures 
will be uniaxially anisotropic with 
the optical axis directed along the polarization vector,
$\vc{E}_{UV}$.
Experimentally, this is, however, not the case.  
For example, constraints imposed by a medium may suppress
out-of-plane reorientation of the azobenzene chromophores
giving rise to the structures with 
strongly preferred in-plane alignment~\cite{Kis:epj:2001}.
Another symmetry breaking effect induced by polymeric environment 
is that the photoinduced orientational structures can be
biaxial~\cite{Wies:1992,Buff:1998,Kis:cond:2001,Kis:epj:2001,Kis:jpcm:2002,Kis:pre:2003}
(a recent review of medium effects on photochemical processes
can be found in~\cite{Ramam:inbk:2005}).

It was recently found that,
similar to the polymer systems, 
the long-term stable POA in the films
containing photochemically stable azo-dye
structures (azobenzene sulfuric dyes)
is characterized by the biaxial photoinduced structures with favored 
in-plane alignment~\cite{Kiselev:idw:2008,Kis:pre:2009}.
Unlike azopolymers, photochromism in  
these films is extremely weak 
so that it is very difficult to unambiguously detect the presence of
a noticeable fraction of \cis\ isomers. 

Over the last decade the photoaligning properties of
such azo-dye (SD1) films has been the subject of intense
studies~\cite{Chigrin:bk:2008,Chig:lc:2002,Chig:pre:2003,Kis:pre2:2005}. 
It was found that, owing to high degree of the photoinduced ordering, 
these films used as aligning substrates are characterized by 
the anchoring energy strengths comparable to the rubbed polyimide films.
For these materials,  the voltage holding ratio and thermal stability
of the alignment turned out to be high. 
The azo-dye films are thus promising materials for applications in
liquid crystal devices. 

The anchoring characteristics 
of the azo-dye films such as the polar and azimuthal anchoring energies 
are strongly influenced by the photoinduced ordering.
In particular, the easy axis is dictated by 
the polarization of the pumping LPUV light,
whereas the azimuthal and polar anchoring
strengths may depend on a number of the governing parameters such as
the wavelength and the irradiation dose~\cite{Kis:pre2:2005}.  

In a LC cell with the initially irradiated layer,
the easy axis may rotate
under the action of secondary irradiation with 
reorienting LP light which polarization
differs from the one used to prepare the aligning layer.
This effect~---~the so-called 
\textit{light-induced reorientation of the easy 
axis}~---~is governed by the \textit{photoinduced reordering} of azo-dye molecules
triggered by the reorienting light and
may be of considerable interest 
for applications such as LC rewritable devices~\cite{Chig:jjap:2008}.
Note that slow reorientation of the easy axis (the light-induced easy axis gliding) 
on the photosensitive layers prepared using 
the PA technique 
was originally observed on poly-(vinyl)-alcohol (PVA) coatings
with embedded azo-dye molecules~\cite{Vorflusev:apl:1997}.
For the azo-dye films, similar results  were reported in 
Ref.~\cite{Pasechnik:lc:2006}.
 
In this paper the kinetics of photoinduced reordering
underlying the light-induced easy axis reorientation on the azo-dye films 
will be our primary concern.
In particular, we 
apply a generalized version of the theoretical analysis presented 
in Ref.~\cite{Kis:pre:2009} 
to examine how
the initial ordering of azo-dye molecules and the characteristics of
reorienting light affect the regime of kinetics and the photosteady states.
 
There are two phenomenological models formulated
in Ref.~\cite{Kis:pre:2009} 
by using different theoretical approaches:
(a)~the two-state model is based on 
the above discussed mechanism of photoisomerization with 
short living excited \cis\ state which is characterized by 
weak photochromism and
negligibly small fraction of \cis\ molecules 
that 
rapidly decays after switching off irradiation;
and 
(b)~the diffusion model describes the light-induced reorientation of
azo-dye molecules as rotational Brownian motion 
governed by the light intensity-dependent mean-field potential.
It turned out that predictions of the models are quite similar
and they both were successfully used to interpret the experimental
data. 
This equivalence points to the fact that,
in the limiting case of short-living \cis\ states,
the photoisomerization cycles with orientation dependent
isomerization rates may generally be viewed as a random walk on the sphere
described by the Fokker-Planck equation for rotational diffusion.
For our purposes, the latter will be conveniently employed
to perform the theoretical analysis and to model the kinetics
of reordering. 

The layout of the paper is as follows.
In Sec.~\ref{sec:model}
we briefly recapitulate the theory~\cite{Kis:pre:2009} 
introducing the diffusion models
where
the process of photoinduced reordering
is treated as a rotational diffusion
in the effective mean-field potential.
Then, in Sec.~\ref{subsec:azim-angle}, 
we discuss the regime of purely in-plane reorientation
and analyze the general properties of
the corresponding two dimensional (2D) diffusion model.
In Sec.~\ref{sec:kinetics}
these results are used to study bifurcations of the photosteady states
and the kinetics of photoinduced reordering 
depending on the initial ordering, the light intensity, $I_{UV}$ 
and  the \textit{polarization azimuth}, $\alpha_p$, 
that characterizes orientation of the polarization vector of 
reorienting LPUV light, $\mathbf{E}_{UV}$.
In Sec.~\ref{sec:experiment}
we apply the model to interpret
the experimental data 
on the light-induced azimuthal gliding of the easy axis
on the photoaligned SD1 substrate
measured as part of the investigation into
effects of the polarization azimuth
in the dynamics of the electrically assisted light-induced 
gliding~\cite{Dubtsov:apl:2012}.
Finally, in Sec.~\ref{sec:conclusion} we draw the results
together and make some concluding remarks. 

\begin{figure*}[!tbh]
\centering
   \resizebox{130mm}{!}{\includegraphics*{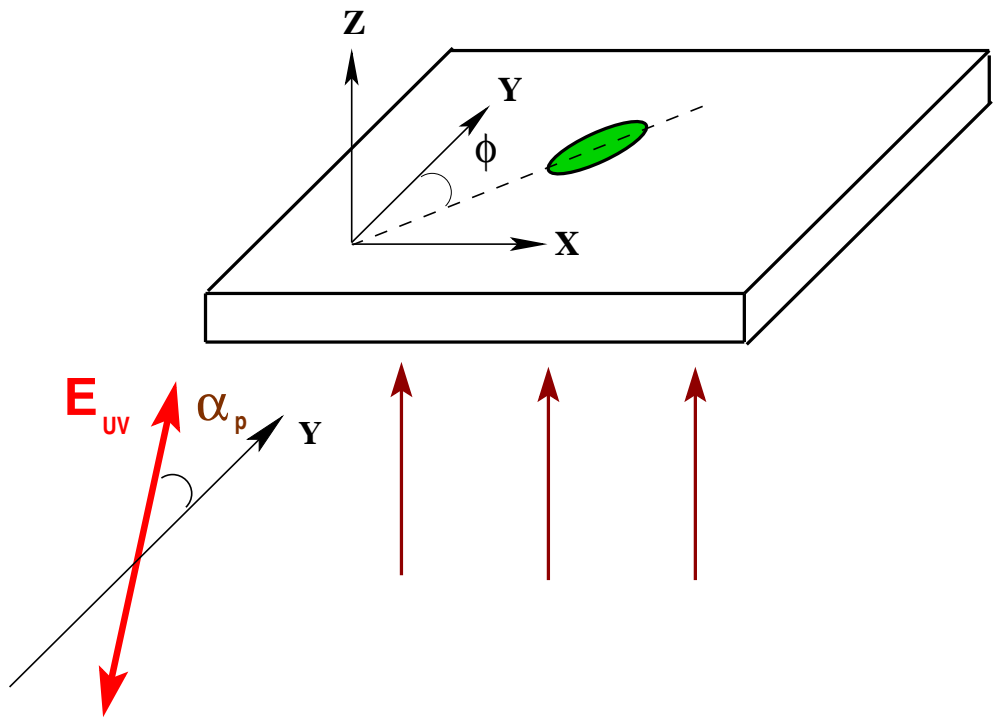}}
\caption{%
(Color online)
Frame of reference:
the $z$ axis is
normal to the substrate and the polarization vector of the
activating light makes the angle $\alpha_p$ with the $y$ axis.  
}
\label{fig:frame}
\end{figure*}

\section{Model}
\label{sec:model}

In this section our starting point is
the approach of Ref.~\cite{Kis:pre:2009}
in which the light-induced reorientation
of azo-dye molecules is described 
as rotational Brownian motion
governed by the effective mean-field potential.

Following Ref.~\cite{Kis:pre:2009}, 
we assume that azo-dye molecules are
cylindrically symmetric, so that
orientation of a molecule 
can be specified by the unit vector, 
$\uvc{u}=(\sin\theta\cos\phi,\sin\theta\sin\phi,\cos\theta)$,
directed along the long molecular axis.
Quadrupolar orientational ordering of such molecules 
is characterized using
the  traceless symmetric second-rank tensor~\cite{Gennes:bk:1993}
\begin{equation}
  \label{eq:Q-def}
  \vc{Q}(\uvc{u})=(3\,\uvc{u}\otimes\uvc{u}-\vc{I}_3)/2,
\end{equation}
where $\vc{I}_n$ is the $n\times n$ identity matrix.
The dyadic~\eqref{eq:Q-def} can be averaged 
over orientation of molecules with 
the normalized \textit{angular distribution function} 
$f(\uvc{u},t)$ to yield
the \emph{order parameter tensor} $\vc{S}$ 
\begin{align}
  \label{eq:avr-Q}
  \avr{\vc{Q}}=\int \vc{Q}(\uvc{u}) f (\uvc{u},t)\,\dd \uvc{u}=
\vc{S},
\end{align}
where $\dd\uvc{u}\equiv\sin\theta\dd\theta\dd\phi$.

\subsection{Rotational mean-field Fokker-Planck equation}
\label{subsec:mf-planck-eq}

The key assumption taken in 
the diffusion models
formulated in Ref.~\cite{Kis:pre:2009}
is that 
the angular distribution function,
$f(\uvc{u},t)$, of rodlike azo-dye molecules
must satisfy the rotational free energy Fokker-Planck
(FP) equation
\begin{align}
  \label{eq:mf_ang_FP_gen}
    \pdrs{t} f =
- \mathcal{L}_{i} 
D_{ij}^{(\mathrm{rot})}
\left\{
f\,\mathcal{L}_{j}\,\vdr{F[f]}{f}
\right\}
\equiv
- \vc{L}
\cdot 
\mvc{D}_{\mathrm{rot}}
\left\{
f\cdot\vc{L}\,\vdr{F[f]}{f}
\right\},
\end{align}
where $D_{ij}^{(\mathrm{rot})}$ is the \textit{rotational diffusion tensor}
and $\vc{L}$ is the angular momentum 
operator that can be expressed in terms of
of the azimuthal and zenithal (polar) angles,
$\phi$ and $\theta$, as follows
\begin{align}
  \label{eq:L_spher}
i\,\vc{L}=[ \vc{r}\times\bnbl ]=
\uvc{e}_{\phi}\, \pdrs{\theta} - [\sin\theta]^{-1} \uvc{e}_{\theta}\, \pdrs{\phi},
\end{align}
where
$\uvc{e}_{\theta}=
(\cos\theta\,\cos\phi,
\cos\theta\,\sin\phi,-\sin\theta)$
and 
$\uvc{e}_{\phi}=
(-\sin\phi,\cos\phi,0)$.

When 
the effective free energy functional
is a sum of the \textit{effective internal energy}, $U[f]$, 
and the Boltzmann entropy,
$\avr{\ln f}$,
$F[f]=U[f]+\avr{\ln f}$,
the free energy FP ~\eqref{eq:mf_ang_FP_gen}
can be recast into the form of
the mean-field FP equation
\begin{align}
   \pdrs{t} f =
- \vc{L}
\cdot 
\mvc{D}_{\mathrm{rot}}
\left\{
\vc{L}\,f
+
f\cdot\vc{L}\,V
\right\},
\quad
V=\vdr{U}{f}.
  \label{eq:mf_ang_FP}
\end{align}
describing the rotational diffusion governed by 
the \textit{effective mean-field potential},
$V$.

\subsubsection{Effective mean-field potential}
\label{subsubsec:mf-poten}

In the lowest order approximation
based on
the truncated expansion for 
the internal energy functional $U[f]$
retaining one-particle (linear) and two-particle (quadratic)
terms,
the effective potential, $V$, is given by
\begin{align}
&
V(\uvc{u})
=\vdr{U}{f(\uvc{u})}=
U_1(\uvc{u})
+
\int
U_2(\uvc{u},\uvc{u}^{\,\prime}) f(\uvc{u}^{\,\prime})\,
\dd\uvc{u}^{\,\prime},
  \label{eq:V_quadr}  
\end{align}
where $U_1(\uvc{u})$ is the external field potential 
and the last term proportional to the symmetrized two-particle kernel,
 $U_2(\uvc{u},\uvc{u}^{\,\prime})$,
represents the contribution coming from 
the intermolecular interactions.

The one-particle part of the effective potential~\eqref{eq:V_quadr}
can be written as a sum of  the light-induced contribution
\begin{align}
\label{eq:U_I}
U_{I}(\uvc{u})=u_{I}\, \cnj{\vc{E}}_{UV}\cdot \vc{Q}(\uvc{u})\cdot \vc{E}_{UV},
\end{align}
where an asterisk indicates complex conjugation,
that comes from the interaction of azo-molecules with the reorienting UV
light and the surface-induced potential
\begin{align}
\label{eq:U_s}
U_s(\uvc{u}) = u_{s}\, \uvc{z}\cdot\vc{Q}(\uvc{u})\cdot\uvc{z}
=u_{s}\, Q_{zz}(\uvc{u})
\end{align}
that takes into account conditions at the bounding
surfaces of the azo-dye layer.
So, for 
the two-particle interaction taken in the Maier-Saupe form 
\begin{align}
  \label{eq:U_2}
  U_2(\uvc{u}_1,\uvc{u}_2)= u_2\, 
\Tr[
\vc{Q}(\uvc{u}_1)\cdot \vc{Q}(\uvc{u}_2)
]=u_2\, Q_{ij}(\uvc{u}_1)Q_{ij}(\uvc{u}_2),
\end{align}
we obtain the following expression for the effective potential
of azo-dye molecules: 
\begin{align}
  \label{eq:V_u}
  V(\uvc{u})=
u_{I} I_{UV}\, \uvc{e}_{p}\cdot \vc{Q}(\uvc{u})\cdot \uvc{e}_{p}
+
u_{s}\, Q_{zz}(\uvc{u})+
  u_2 S_{ij}\, Q_{ij}(\uvc{u}),
\end{align}
where $I_{UV}$ is the intensity of the  reorienting
UV light which is assumed to be linearly polarized 
along the unit polarization vector
$\uvc{e}_{p}$  
\begin{align}
  \label{eq:alpha-p}
  \uvc{e}_{p}=\mvc{R}_{\ind{rot}}(\alpha_p)\,\uvc{y},
\quad
\mvc{R}_{\ind{rot}}(\alpha_p)=
\begin{pmatrix}
  \cos\alpha_p &-\sin\alpha_p\\
\sin\alpha_p& \cos\alpha_p
\end{pmatrix}
\end{align}
that makes the angle $\alpha_p$ with 
the $y$ axis directed along the polarization vector
of the initial irradiation: $\vc{E}_{\ind{ini}}\parallel \uvc{y}$
(see Fig.~\ref{fig:frame}).

Note that, in the limiting case of non-interacting molecules  with $U_2=0$,
the rotational mean-field FP equation~\eqref{eq:mf_ang_FP}
reduces to the well-known linear rotational diffusion equation 
that has been widely used to study
a large variety of problems based on the rotational diffusion model
for the rotational motion of molecules in the presence
of external fields.
These include
dielectric and Kerr effect relaxation of polar 
liquids~\cite{Lauritz:advmrel:1973,Coffey:jcp:1993,Dejardin:jcp:1993,Dejardin:pre:2000,Kalmykov:jcp:1991,Felderhof:pre:2002,
Kalmykov:pre:2001,Kalmykov:jcp:2007},
rotational diffusion of a probe molecule dissolved in a 
liquid crystal 
phase~\cite{Nordio:jcp:1971,Zannoni:jcp:1991,Luckh:molph:1972,
Zannoni:jcp:1993,Ferrarini:inbk:1994,Zannoni:jcp:2000}
and the molecular reorientation in liquid crystal 
phases~\cite{Nordio:inbk:1979,Moro:mclc:1984,Dozov:pra:1987,Fontana:pra:1986}.
  
\subsubsection{Steady states}
\label{subsubsec:steady-states}
 
The equilibrium angular distribution can generally be obtained as a 
stationary solution to the FP equation~\eqref{eq:mf_ang_FP}.
It can be readily checked that the Boltzmann distribution 
\begin{align}
  \label{eq:st_dist_gen}
  f_{\,\st}(\uvc{u})= Z_{\,\st}^{-1}\,
\exp[-V(\uvc{u})],
\quad
Z_{\st}=\int\exp[-V(\uvc{u})]\dd\uvc{u}
\end{align}
determined by 
the effective potential gives the stationary solution.
Note that 
the energy scale is determined by 
the temperature factor $k_{B} T$ incorporated into 
the rotation diffusion tensor, so that
the internal energy, $U[f]$, and the potential, $V$, 
are both dimensionless.

When the FP equation is linear,
the stationary distribution~\eqref{eq:st_dist_gen}
describing the equilibrium state is unique.
By contrast, this is no longer the case when
the effective potential~\eqref{eq:V_u}
depends on the elements of the averaged 
orientational order parameter tensor~\eqref{eq:avr-Q}: 
$V(\uvc{u})=V(\uvc{u}|\mvc{S})$. 
In this case,  the components of the order parameter
tensor in the stationary state, $S_{ij}=S_{ij}^{(\st)}$,
can be found from
the self-consistency equation
\begin{align}
  \label{eq:self_cons_gen}
  S_{ij}=
\int Q_{ij}(\uvc{u})\, f_{\,\st}(\uvc{u}|\mvc{S})\,
\dd\uvc{u}.
\end{align}
which generally possesses
several solutions  
representing multiple local extrema
(stationary points) of the free energy
\begin{align}
  \label{eq:F_st_gen}
  F[f_{\,\st}]\equiv F_{\,\st}(\mvc{S})
=
-\frac{u_2}{2}\, S_{ij} S_{ij} - \ln Z_{\,\st}(\mvc{S}).
\end{align}

Following the line of reasoning presented in 
Ref.~\cite{Frank:bk:2005} and 
using the effective free energy
as the Lyapunov functional, it is not difficult to prove
the $H$ theorem for nonlinear FP equations
of the form~\eqref{eq:mf_ang_FP_gen}.
It follows that, in the long time limit, 
all transient solutions converge to stationary ones. 
So, each stable stationary distribution
is characterized by  the basin of attraction giving
orientational states (angular distributions)
that evolve in time approaching 
the stationary distribution.

\subsection{Two-dimensional model of purely in-plane reorientation}
\label{subsec:azim-angle}

In the previous section 
our model has been formulated as the free energy FP equation~\eqref{eq:mf_ang_FP}
describing rotational diffusion of azo-dye molecules
governed by the effective mean field potential~\eqref{eq:V_u}.
In this section,
we concentrate on
the limiting case of 
purely in-plane reorientation
and deal with
the simplified two-dimensional model 
\begin{align}
    \pdrs{\tau} f =
&
\pdrs{\phi}
\Bigl[
\pdrs{\phi} f
+f \pdrs{\phi} V
\Bigr]
= \pdrs{\phi}^2 f +\frac{1}{2}\,
\Bigl[\,
\pdrs{\phi}^2 (f V)
\nonumber
\\
&
+ f \pdrs{\phi}^2 V-
V \pdrs{\phi}^2 f
\Bigr],
\quad
\tau = D_z^{(\mathrm{rot})} t.
 \label{eq:mf_FP_phi}  
\end{align}
can be derived by assuming that
the out-of-plane component of the unit vector
$\uvc{u}$ describing orientation of
the azo-dye molecules is suppressed
and, as is shown in Fig.~\ref{fig:frame}, 
$\uvc{u}=(\sin\phi,\cos\phi,0)$.
It implies that, 
similar to the model 
of a single axis rotator with two
equivalent sites~\cite{Lauritz:advmrel:1973,Coffey:jcp:1993},
the molecules are constrained
to be parallel to the substrate plane
(the $x$-$y$ plane). 

\subsubsection{Potential}
\label{subsubsec:potential-2D}

When the out-of-plane
reorientation is completely suppressed
and $\uvc{u}=(\sin\phi,\cos\phi,0)$,  
the order parameter dyadic~\eqref{eq:Q-def}
simplifies as follows
\begin{align}
  \label{eq:Q-2D}
  \mvc{Q}(\phi)
=
\begin{pmatrix}
  \mvc{Q}_2(\phi) & \mvc{0}\\
\mvc{0} & -1/2
\end{pmatrix}
\end{align}
where $\mvc{Q}_2$ is the two-dimensional in-plane part of the tensor.
The \textit{in-plane order parameter tensor} is given by the $2\times 2$ block-matrix
\begin{align}
  \label{eq:Q2-def}
  4 \mvc{Q}_2(\phi)=\mvc{I}_2-3\cos(2\phi)\,\bs{\sigma}_3+3\sin(2\phi)\,\bs{\sigma}_1
\end{align}
expressed in terms of the Pauli matrices:
$\bs{\sigma}_3=\begin{pmatrix} 1& 0\\ 0 & -1 \end{pmatrix}$
and $\bs{\sigma}_1=\begin{pmatrix} 0& 1\\ 1 & 0 \end{pmatrix}$.
We can now substitute the expression for the order parameter
tensor~\eqref{eq:Q-2D}
into the effective potential~\eqref{eq:V_u}
and use the algebraic identity
\begin{align}
  \label{eq:Q2-rel}
  \mvc{R}_{\ind{rot}}(-\alpha) \mvc{Q}_2(\phi)
  \mvc{R}_{\ind{rot}}(\alpha)
=\mvc{Q}_2(\phi+\alpha)
\end{align}
to derive the angular dependent part of
the potential in the following form:
\begin{subequations}
\label{eq:V_azim}
\begin{align}
&
  V=\bigl(
\cnj{v} \exp[ 2 i \phi]
+ v \exp[ -2 i \phi]
\bigr)/2,
  \label{eq:V_complex}
\\
&
v=v_1\exp[ -2 i \alpha_p]+v_2\avr{\exp[ 2 i \phi]},
  \label{eq:V_coeff}
\\
&
v_1\equiv 3 u_{I} I_{UV}/4,\quad
v_2=9 u_2/8,
  \label{eq:v1_v2}
\end{align}
\end{subequations}
where 
$\displaystyle
\avr{\ldots}=\int_{0}^{2\pi}\ldots f \dd\phi$;
$v_1$ and
$v_2$ are the \textit{photoexcitation} and 
\textit{intermolecular interaction} parameters,
respectively.

\subsubsection{Fourier harmonics and order parameters}
\label{subsubsec:harmonics}

Our next step is to obtain
the system of equations for the averaged complex harmonics,
$h_n(\tau)=\avr{\exp[ i n\phi]}(\tau)$,
that  are proportional to the Fourier coefficients
of the distribution function, $f(\phi,\tau)$
\begin{align}
  \label{eq:Fourier-f}
  f(\phi,\tau)=(2\pi)^{-1} \sum_{n=-\infty}^{\infty} h_n(\tau) \exp[ -i n\phi].
\end{align}
To this end, we integrate the FP equation~\eqref{eq:mf_FP_phi}
multiplied by $\exp[i n\phi]$ over the azimuthal angle.
The resulting system reads
\begin{align}
  \label{eq:sys_h_n}
  \pdrs{\tau} h_n(\tau)= - n^2 h_n + 
n\,\bigl\{
\cnj{v}(\alpha_p,h_2) h_{n+2}(\tau)- v(\alpha_p,h_2) h_{n-2}(\tau)
\bigr\}, 
\end{align}
where $v(\alpha_p,h_2)\equiv v=v_1\exp[ -2 i \alpha_p]+v_2 h_2$.

When $f(\phi+\pi)=f(\phi)$, 
the odd numbered harmonics vanish,
$h_{2 k +1}=0$.
For the even numbered harmonics,
$\rho_k\equiv h_{2 k}$,
the system~\eqref{eq:sys_h_n} 
can be conveniently recast into the form 
\begin{align}
& 
   \pdrs{\tau} \rho_k= - 4 k^2 \rho_k + 2 k\,(\cnj{v} \rho_{k+1}-v \rho_{k-1}),
\quad
k=1,2,\ldots
\label{eq:sys_p_n}
\\
& 
\rho_{k}=p_k+i q_k =\avr{\cos (2 k \phi)}+ i \avr{\sin(2 k \phi)}\equiv h_{2k},
\quad
\rho_{0}= 1,
\end{align}
where $\rho_1=p_1+i q_1=\avr{\exp [2i \phi]}$ is 
the complex \textit{order parameter harmonics}.

For the in-plane order parameter tensor~\eqref{eq:Q2-def} 
averaged over the azimuthal angle
\begin{align}
&
  \label{eq:S_2-def}
  \avr{\mvc{Q}_2}=\mvc{S}_2
=\frac{1}{4}
\begin{pmatrix}
  1-3 p_1 & q_1\\
q_1 & 1+ 3 p_1
\end{pmatrix}
=
\mvc{R}_{\ind{rot}}(\alpha_e)
\begin{pmatrix}
  s_{+} & 0\\
0 & s_{-}
\end{pmatrix}
\mvc{R}_{\ind{rot}}(-\alpha_e),
\\
&
\label{eq:S_2-eigval}
s_{\pm}=(1\pm 3 |\rho_1|)/4,
\quad
2 \alpha_e=\arg(-\cnj{\rho_1}),
\end{align}
the order parameter harmonics, $\rho_1$,  defines 
the principal values (eigenvalues), $s_{+}$ and $s_{-}$,
and 
the eigenvector,
$(\cos\alpha_e,\sin\alpha_e)$, 
for the largest eigenvalue $s_{+}$
giving the in-plane principal axis:
$\uvc{n}_e=(\cos\alpha_e,\sin\alpha_e,0)$.
So, the tensor~\eqref{eq:S_2-def}
can be rewritten in the form of the in-plane
order parameter dyadic: 
\begin{align}
  \label{eq:S_2_dyad}
  \mvc{S}_2=
s_{-}\mvc{I}_2
+(s_{+}-s_{-})
\uvc{n}_{e}\otimes\uvc{n}_{e}.
\end{align}

Temporal evolution of the in-plane order parameter tensor
is determined by the order parameter harmonics, $\rho_1(t)$,
that can be computed as a function of time by solving 
the initial value problem for
the system~\eqref{eq:sys_p_n} 
with 
the initial values of the harmonics 
$\rho_k(0)=\rho_k^{(\ind{ini})}$
representing the initial angular distribution 
$f(\phi,0)=f_{\ind{ini}}(\phi)$.

\subsubsection{Symmetry}
\label{subsubsec:symmetry}

It is not difficult to check that
the system~\eqref{eq:sys_h_n}
is invariant under the symmetry transformation
\begin{align}
  \label{eq:symmetry}
  h_n\to \tilde{h}_n=\exp[i n \alpha]h_n,\quad
v\to \tilde{v}=\exp[2 i\alpha] v=v(\alpha_p-\alpha,\tilde{h}_2).
\end{align}
An important consequence of 
these  relations
concerns the special case where the reorienting light
is switched off and $v_1=0$.
It can be readily seen that, 
at $v=v_2 \rho_1$,
a set of the harmonics, $\{\rho_k\}_k$,
and that of the transformed harmonics $\{\exp[2ik\alpha] \rho_k\}_k$  
both satisfy the system of equations~\eqref{eq:sys_p_n}.
In particular, at $I_{UV}=0$,
the stationary angular distribution, $f_{\st}(\phi)$,
and the shifted (rotated) one,  $f_{\st}(\phi-\alpha)$,
both represent steady states.

By using the symmetry relations~\eqref{eq:symmetry}
dynamical equations~\eqref{eq:sys_p_n}
can be conveniently changed into the system with 
the zero polarization azimuth, $\alpha_p=0$.
More precisely, from Eq.~\eqref{eq:symmetry},
it follows that
the transformed harmonics
\begin{align}
  \label{eq:tilde_rho}
 \tilde{\rho}_k=\exp[2ik\alpha_p]\rho_k  
\end{align}
satisfy the system~\eqref{eq:sys_p_n} with  the
coupling coefficient $v(\alpha_p,\rho)$ 
and the initial distribution $f_{\ind{ini}}(\phi)$
replaced by $\tilde{v}=v(0,\tilde{\rho}_1)$
and $\tilde{f}_{\ind{ini}}=f_{\ind{ini}}(\phi-\alpha_p)$,
respectively. 
Geometrically, such transformation can be described as 
a transition to the frame of reference where the $y$
axis is directed along the
polarization vector of reorienting light.

So, in the subsequent section, we shall restrict
our considerations to the special
case of the system with the coupling
coefficient 
\begin{align}
  \label{eq:tilde_v}
 \tilde{v}=v_1+v_2\tilde{\rho}_1
\end{align}
taken at $\alpha_p=0$.
In this case, the polarization
azimuth, $\alpha_p$, enters 
the initial data, 
$\tilde{\rho}_k^{(\ind{ini})}=\exp[2ik\alpha_p]\rho_k^{(\ind{ini})}$, 
and the formula
\begin{align}
  \label{eq:alph_e}
  \alpha_e=\alpha_p+\frac{1}{2} \arg(-\cnj{\tilde{\rho}_1})
\end{align}
gives the azimuthal angle of the
principal axis~\eqref{eq:S_2-eigval}
expressed in terms of the order
parameter harmonics $\tilde{\rho}_1$.

\begin{figure*}[!tbh]
\centering
\resizebox{150mm}{!}{\includegraphics*{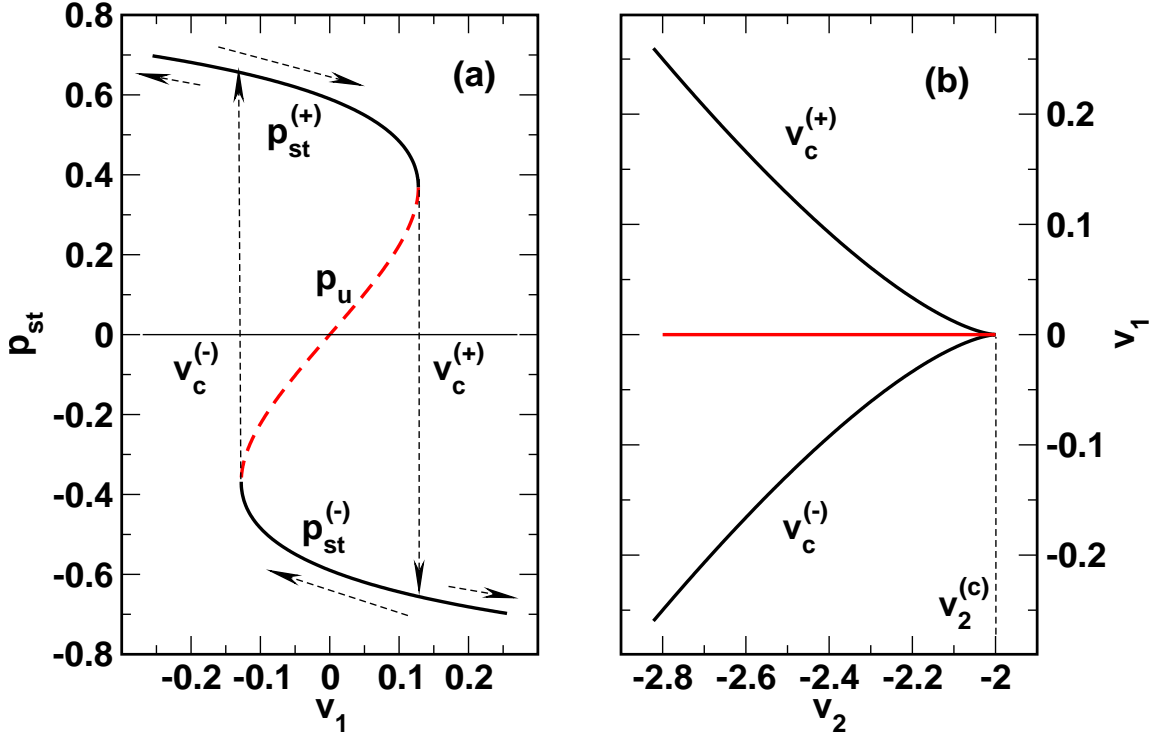}}
\caption{%
(Color online)
(a)~Steady state order parameter harmonics $p_{\st}$
as a function of the photoexcitation parameter $v_{1}$ at $v_2=-2.5$;
(b)~bifurcation curves 
in the $v_{2}$-$v_1$ plane
are typical of the cusp catastrophe 
with the cusp singularity located at $(-2,0)$.
}
\label{fig:bifur}
\end{figure*}

\begin{figure*}[!tbh]
\centering
\resizebox{150mm}{!}{\includegraphics*{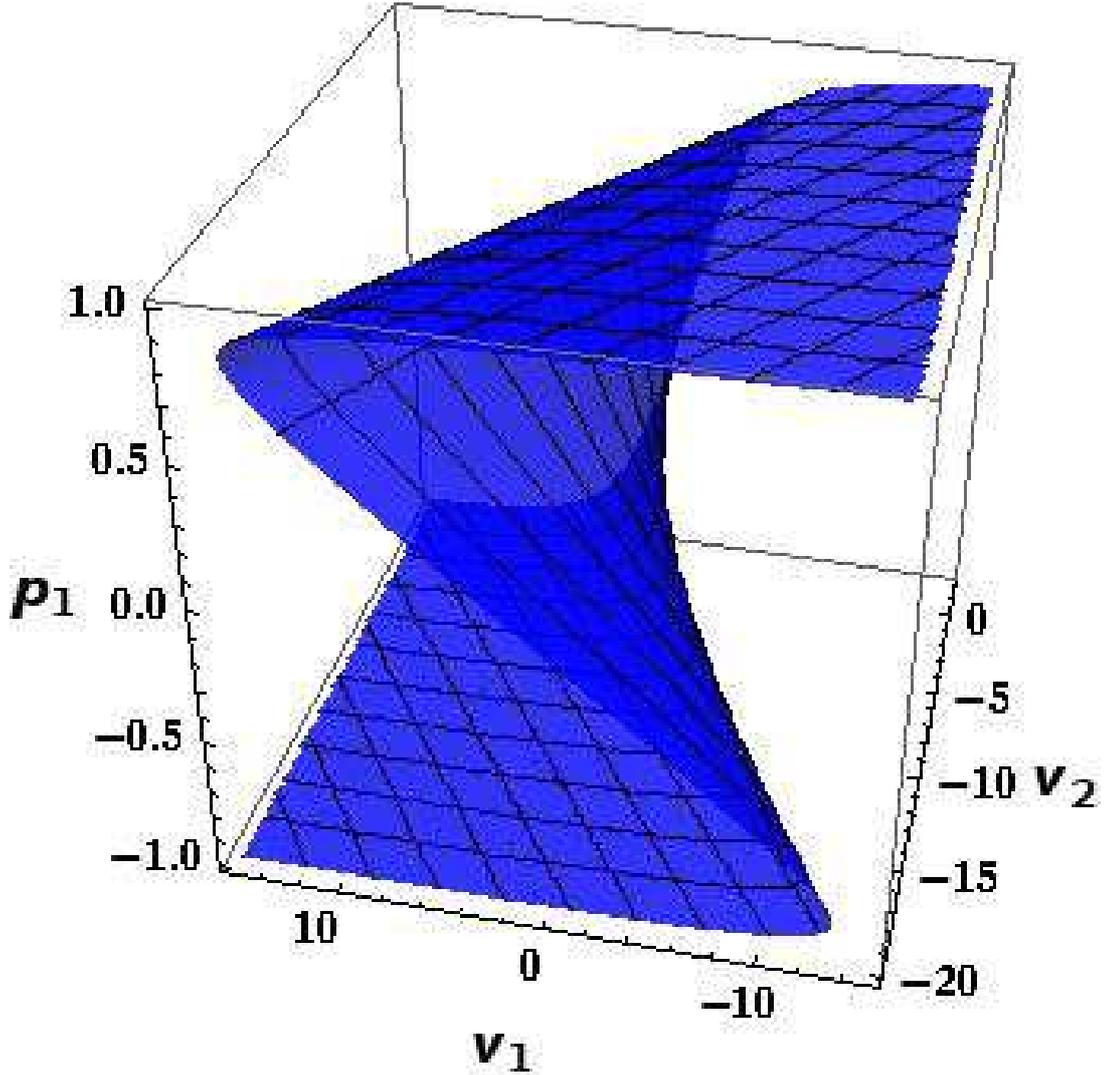}}
\caption{%
(Color online)
Bifurcation diagram as
the cusp surface in 
the three dimensional 
$(v_{1},v_{2},p_{1})$ space.
}
\label{fig:cusp}
\end{figure*}

\section{Kinetics of photoinduced in-plane  reordering}
\label{sec:kinetics}

\subsection{Bifurcations of equilibria}
\label{subsec:steady-state-bifur}

The expression
for the stationary distributions 
\begin{align}
  \label{eq:f_st_azim}
  f_{\st}=Z_{\st}^{-1}\exp[-V]=Z_{\st}^{-1}\exp[-|\tilde{v}|\cos (2\phi-\tilde{\alpha})],
\quad
Z_{\st}=\int_{0}^{2\pi}\exp[-V]\dd\phi
\end{align}
comes from
the general formula~\eqref{eq:st_dist_gen}
and represents the photosteady states
in the two dimensional case with the
potential given in Eq.~\eqref{eq:V_azim}
where the coupling coefficient is changed from
$v$ to
$\tilde{v}= |\tilde{v}|\exp[i\tilde{\alpha}]$.

Equation~\eqref{eq:f_st_azim}
can now be combined with the relation
\begin{align}
  \label{eq:rel_bessel_1}
  \exp[-v\cos 2\phi]=
I_{0}(v)+
2 \sum_{k=1}^{\infty} (-1)^k I_{k}(v)\cos 2k\phi, 
\end{align}
where $I_{k}$ is 
the modified Bessel function of integer order~\cite{Abr},
to derive the stationary state statistical integral
\begin{align}
&
  \label{eq:Z_st_azim}
  Z_{\st}= 2\pi I_{0}(|\tilde{v}|)
\end{align}
and the stationary state
free energy~\eqref{eq:F_st_gen}
\begin{align}
&
  \label{eq:F_st_azim}
F_{\st}(\tilde{v})=v_{2}^{-1}
\Bigl[
-\frac{|\tilde{v}|^2}{2}+v_{1} |\tilde{v}|\cos\tilde{\alpha}
\Bigr] 
- \ln I_{0}(|\tilde{v}|), 
\end{align}
where the additive constant is chosen so as to
have the free energy  vanishing at $\tilde{v}=0$. 

The stationary points of the energy~\eqref{eq:F_st_azim},
$\tilde{v}=\tilde{v}_{\st}=v_1+v_2 \tilde{\rho}_1^{(\st)}$, 
define
the steady state distributions~\eqref{eq:f_st_azim} 
characterized by
the stationary values of the order parameter harmonics,
$\tilde{\rho}_1^{(\st)}=\tilde{p}_1^{(st)}+i \tilde{q}_1^{(\st)}$,
which can also be found as solutions of 
the self-consistency equation~\eqref{eq:self_cons_gen}.
Close inspection of the formula~\eqref{eq:F_st_azim} 
shows that, 
for stationary points,
the steady state coupling (potential strength) coefficient
is real, $\tilde{v}_{\st}=v_1+v_2 \tilde{p}_1^{(st)}\equiv v_1+v_2 p_{\st}$, with $\tilde{q}_1^{(st)}=0$
and $\sin\tilde{\alpha}_{\st}=0$. 
So, we may
closely follow the line of reasoning
presented  in Ref.~\cite{Kis:pre:2009}
to perform the bifurcation analysis. 

According to Ref.~\cite{Kis:pre:2009},
the steady state harmonics
are given
\begin{align}
\tilde{\rho}_k^{(st)}=\tilde{p}_{k}^{(\st)}= (-1)^k I_{k}(\tilde{v}_{\st})/I_{0}(\tilde{v}_{\st}),
  \label{eq:pk_st_azim}  
\end{align}
where $\tilde{v}_{\st}$ satisfies the self-consistency condition 
\begin{align}
  \label{eq:self_cosist_azim}
  p_{\st}=(\tilde{v}_{\st}-v_{1})/v_{2}= - I_{1}(\tilde{v}_{\st})/I_{0}(\tilde{v}_{\st}).
\end{align}
As is evident from the curve depicted in Fig.~\ref{fig:bifur}(a),
the number of solutions
of the self-consistency equation~\eqref{eq:self_cosist_azim}
varies between one and three
depending on the values of the parameters $v_1$ and $v_2$.
Figure~\ref{fig:bifur}(a) shows the order parameter harmonics,
$\tilde{p}_{1}^{(\st)}\equiv p_{\st}$,
plotted in the $v_1$-$p_1$ plane by using 
the parametrization
\begin{align}
  \label{eq:p1_st_v1}
p_{\st}=
\begin{cases}
  p_{1}=p_{1}(\xi)=- I_{1}(\xi)/I_{0}(\xi),\\
v_1=v_1(\xi)=\xi-v_2\, p_{1}(\xi)
\end{cases}
\end{align}
and demonstrates that,
for the intermolecular interaction parameter $v_2=-2.5$
and sufficiently small  photoexcitation parameters,
the steady state free energy~\eqref{eq:F_st_azim} has
two local minima separated by the energy barrier
and thus possesses three stationary points. 
More specifically, the steady states are determined by
the free energy of the double-well potential form
only if the inequalities
\begin{align}
  \label{eq:bimodal}
  v_{2}<v_{2}^{(c)}=-2,
\quad
v_{c}^{(-)}<v_{1}<v_{c}^{(+)}
\end{align}
are satisfied.

The critical values of the parameter $v_1$
depend on the reduced strength of 
intermolecular interaction $v_2$
and can be parametrized as follows
\begin{align}
  \label{eq:vc_st_v2}
v_{c}=
\begin{cases}
v_1=v_1(\xi)=\xi-v_2 p_{1}(\xi),\\
  v_{2}=v_{2}(\xi)=- [\bigl(I_{1}(\xi)/I_{0}(\xi)\bigr)'_{\xi}]^{-1}=\dfrac{1}{1+p_{1}(\xi)/\xi-p_{1}^{2}(\xi)}.
\end{cases}
\end{align}
Geometrically, in the $v_{2}$-$v_{1}$ plane,
 equation~\eqref{eq:vc_st_v2} defines
the bifurcation curves shown in Fig.~\ref{fig:bifur}(b). 
These curves form a bifurcation set 
which is the projection of the cusp surface
\begin{align}
  \label{eq:bifur_surf_param}
S_{B}=
  \begin{cases}
v_1=\xi-\zeta p_{1}(\xi),\\
v_2=\zeta,\\
  p_{1}=- I_{1}(\xi)/I_{0}(\xi)
\end{cases}
\end{align}
representing the bifurcation diagram 
in the three dimensional $(v_1,v_2,p_1)$  
space (see Fig.~\ref{fig:cusp}).
Note that the cusp bifurcation occurs as a canonical model of
a codimension 2 singularity~\cite{Kuznetsov:bk:1998}
and the surface shown in  Fig.~\ref{fig:cusp}
is typical of the cusp catastrophe~\cite{Hale:bk:1991,Hoppen:bk:2000}.

\subsection{Steady-state order parameter tensor}
\label{subsec:steady-state-order}

From the discussion at the end of
Sec.~\ref{subsubsec:symmetry}
and the relation~\eqref{eq:pk_st_azim},
 at $\alpha_p\ne 0$, 
the stationary angular distributions
are defined by the steady state harmonics
\begin{align}
  \label{eq:st-harmonics}
\rho_k^{(\ind{\st})}  =\exp[-2ik\alpha_p]
(-1)^k I_{k}(\tilde{v}_{\st})/I_{0}(\tilde{v}_{\st}),
\end{align}
which are expressed in terms of the steady state
coupling coefficient,
$\tilde{v}_{\st}=v_1+ v_2 p_{\st}$,
that meets the self-consistency condition~\eqref{eq:self_cosist_azim}.
For each steady state characterized by the corresponding solution of 
Eq.~\eqref{eq:self_cosist_azim}, $\tilde{v}_{\st}$, 
the principal values of
the in-plane order parameter tensor~\eqref{eq:S_2-def},
$s_{\pm}^{(\st)}$,
can be computed from Eq.~\eqref{eq:S_2-eigval} 
by setting $\rho_1=p_{\st}=(\tilde{v}_{\st}-v_1)/v_2$,
whereas the azimuthal angle, $\alpha_e^{(\st)}$,
characterizing  steady state orientation  of the principal axis
is given by the formula~\eqref{eq:alph_e}
with $\tilde{\rho_1}=p_{\st}$:
\begin{subequations}
 \label{eq:s_alph_st}
\begin{align}
&
  \label{eq:s_st}
s_{\pm}^{(\st)}=(1\pm 3 |p_{\st}|)/4,
\\
&
  \label{eq:alph_st}
  \alpha_e^{(\st)}=\alpha_p+\frac{1}{2} \arg(-p_{\st})
=
\begin{cases}
  \alpha_p\text{ and } \uvc{n}_e^{(\st)}\perp\vc{E}_{UV}, & p_{\st}<0\\
\alpha_p+\pi/2\text{ and } \uvc{n}_e^{(\st)}\parallel\vc{E}_{UV}, &p_{\st}>0
\end{cases}.
\end{align}
\end{subequations}

\begin{figure}[!htb]
\centering
\subfloat[Order parameter harmonics]{
\resizebox{80mm}{!}{\includegraphics*{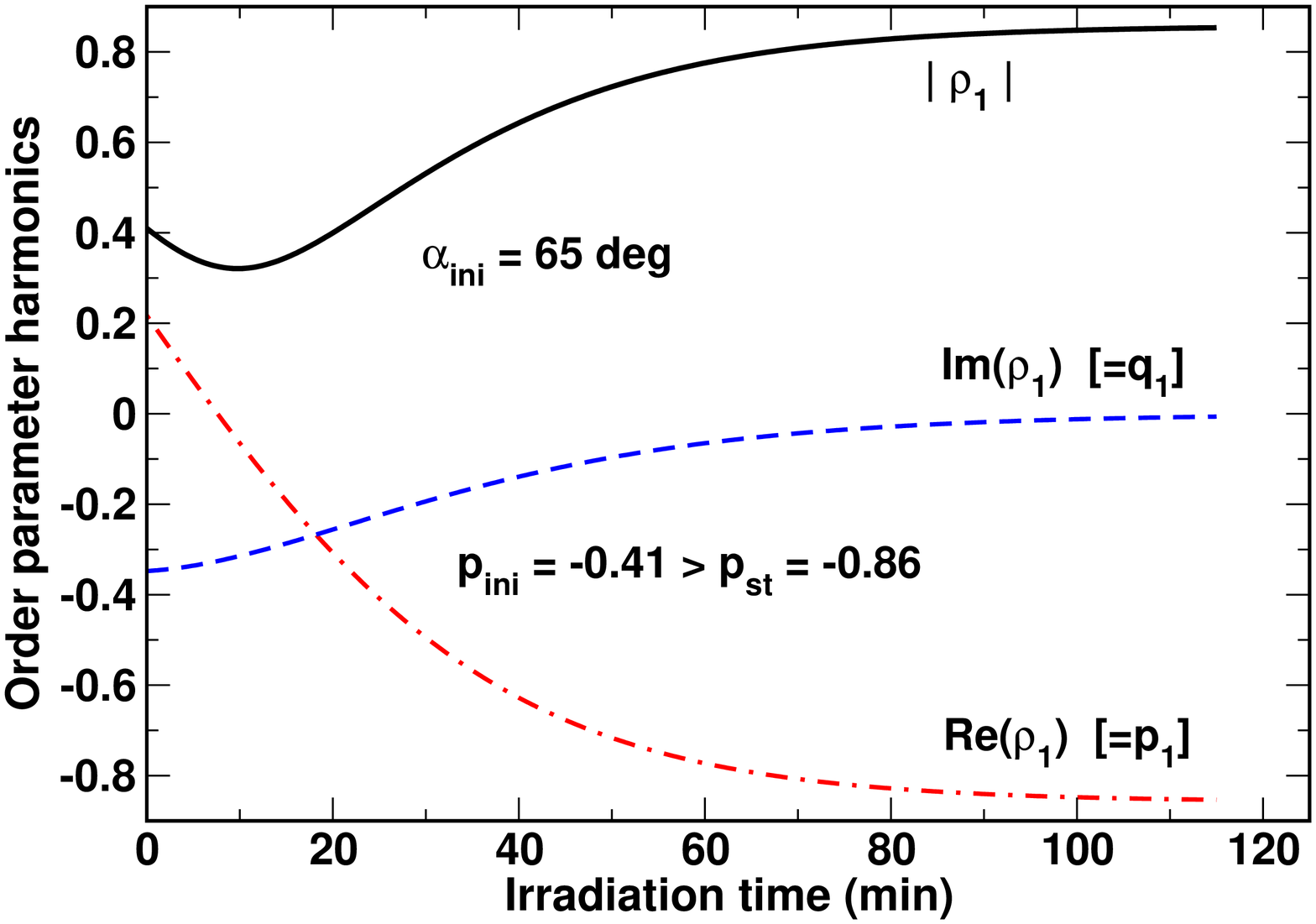}}
\label{fig:rho1-65-mono}
}
\subfloat[Angular distribution function]{
\resizebox{70mm}{!}{\includegraphics*{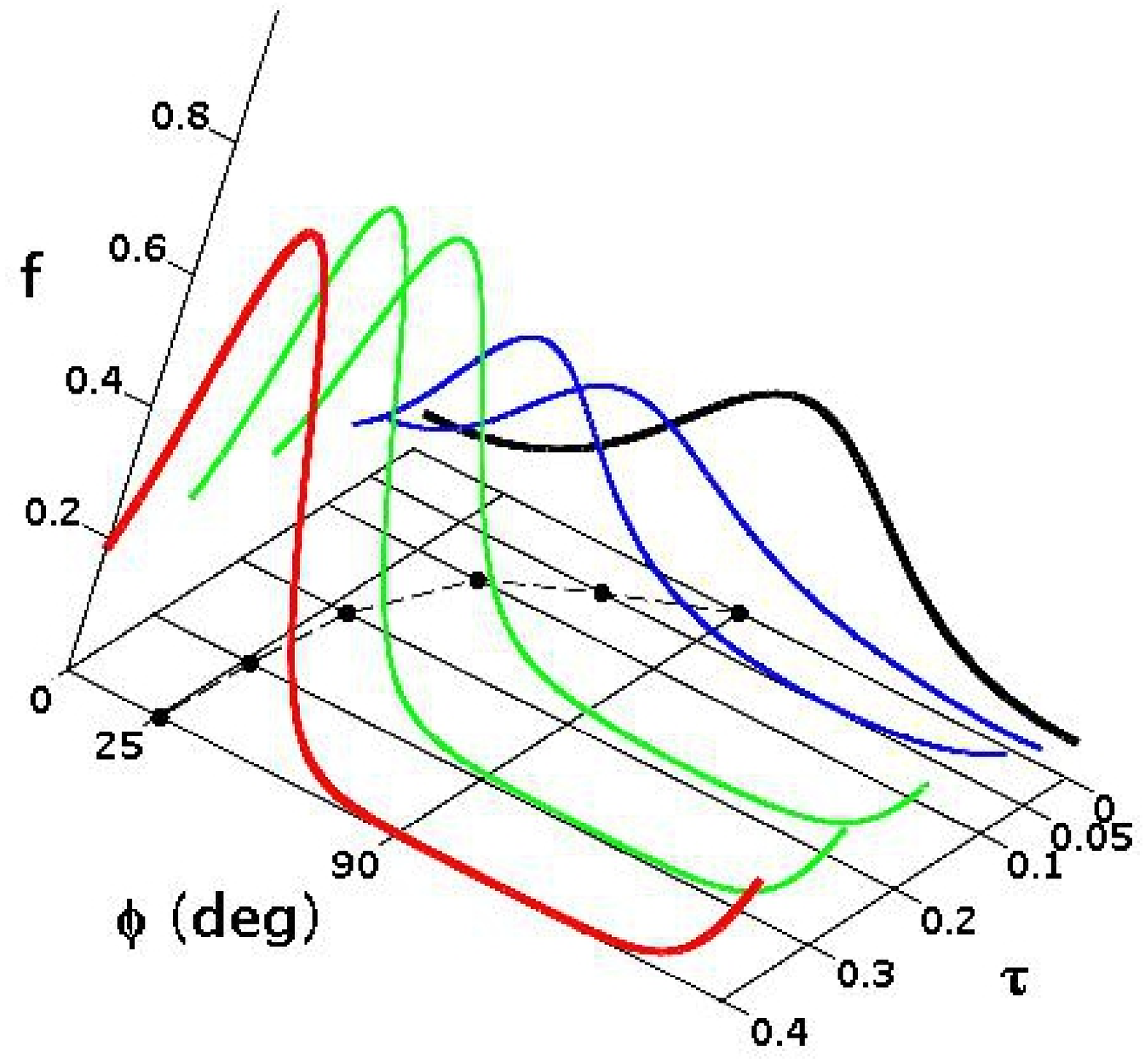}}
\label{fig:distr-65-mono}
}
\caption{%
(a)~Order parameter harmonics,
$\rho_1$,
computed as a function of dimensionless time parameter
$\tau$ by  numerically  solving
the initial value problem for 
the system~\eqref{eq:sys_p_n}
with the initial angular distribution~\eqref{eq:initial-distr}
at $\alpha_{\ind{ini}}=65$~deg
and $p_{\ind{ini}}=-0.41$. 
The photoexcitation and interaction parameters are:
$v_1=3$ and $v_2=-1$.
The stationary state is unique and is characterized by
the steady state harmonics
$p_{\st}=p_{\st}^{(-)}\approx -0.86$.
(b)~Angular dependence of the orientational distribution 
function~\eqref{eq:ang-distr-func}, $f(\phi,\tau)$,
at  different values of the time parameter,
$\tau$. The distribution is $\pi$-periodic 
and solid circles indicate the angles,
$\phi_{\ind{max}}=\pi/2-\alpha_e$, where peaks are located.
}
\label{fig:65-mono}
\end{figure}

An important conclusion to draw from this result is that,
upon reaching the stationary state,
the principal axis reorients along the direction perpendicular (parallel)
to the polarization vector of the reorienting light,
$\uvc{e}_p=(-\sin\alpha_p,\cos\alpha_p,0)$, with $\alpha_e\to \alpha_e^{(\st)}=\alpha_p$
($\alpha_e\to \alpha_e^{(\st)}=\pi/2+\alpha_p$)
when the steady state order parameter harmonics
is negative (positive), $p_{\st}=p_{\st}^{(-)}<0$ 
($p_{\st}=p_{\st}^{(+)}>0$). 
Thus, in the photosaturated regime, 
the orientation of azo-dye molecules turned out to be defined solely by  
the polarization azimuth of the reorienting UV light.

As it has been discussed in Sec.~\ref{subsubsec:symmetry},
when the activating light is switched off and
$I_{UV}=0$ ($v_1=0$), the steady states are 
continuously degenerate
with the azimuthal angle of the principal axis
playing the role of the degeneracy parameter.
The reason for such degeneracy is that
the intermolecular interaction is taken in 
the rotationally invariant 
Maier-Saupe form~\eqref{eq:U_2}. 
The effect of the reorienting linearly polarized light
is to lift the degeneracy
by breaking the rotational symmetry
of the intermolecular interaction.

\begin{figure}[!htb]
\centering
\resizebox{120mm}{!}{\includegraphics*{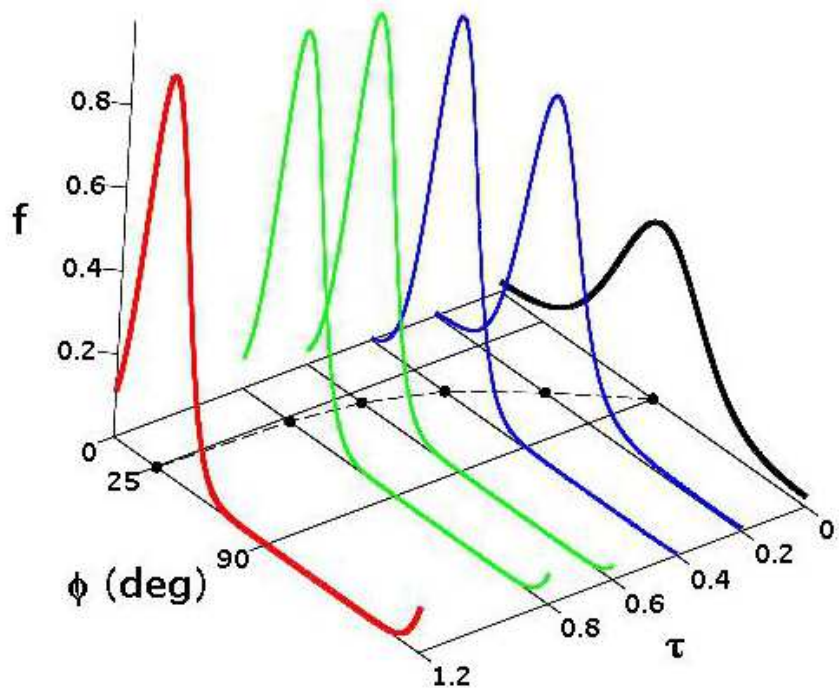}}
\caption{%
Angular dependence of the orientational distribution function,
$f$, within the span of the period
computed for various values of the time parameter,
$\tau=D_{\ind{rot}} t$,  at $\alpha_{\ind{ini}}=65$~deg
and $p_{\ind{ini}}=-0.58$. 
The photoexcitation and interaction parameters are
$v_1=1.3$ and $v_2=-5$, respectively.
}
\label{fig:distr-bstb}
\end{figure}

\subsection{Computational procedure
and bistability effects}
\label{subsec:modeling}
 
According to the results of Ref.~\cite{Kis:pre2:2005},
when the photoaligned azo-dye film
is used as an aligning substrate in a NLC cell,
orientation of the easy axis at the azo-dye substrate
is dictated by the principal axis of the order
parameter tensor~\eqref{eq:S_2-def}.
In particular, under certain conditions,
the difference between 
the angle $\alpha_e$ and the easy axis azimuthal angle
can be negligible.
In this case, the kinetics of the easy axis
can be modeled
by solving the system~\eqref{eq:sys_p_n} 
and computing the angle $\alpha_e$
(see Eq.~\eqref{eq:alph_e})
as a function of time.

\begin{figure}[!htb]
\centering
\resizebox{120mm}{!}{\includegraphics*{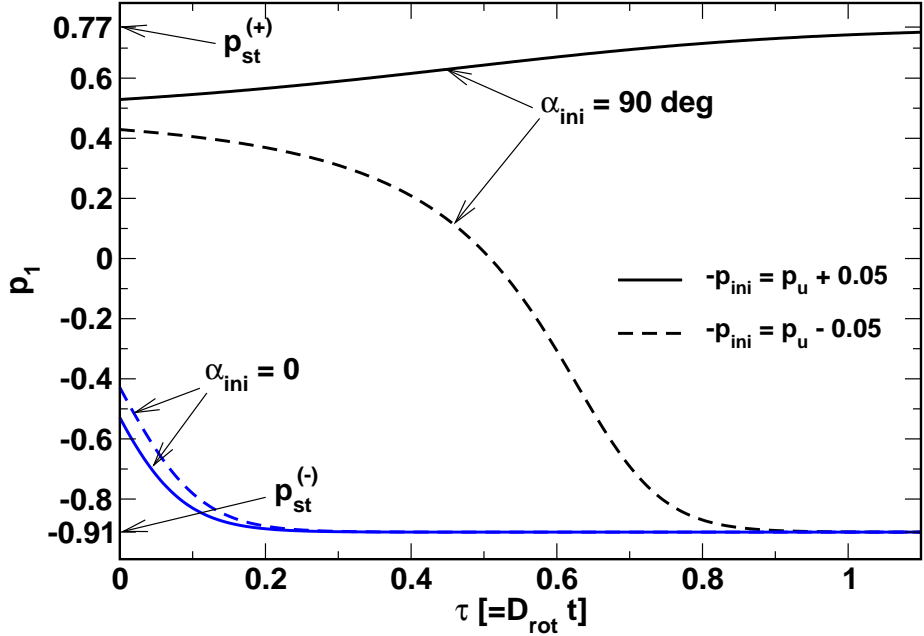}}
\caption{%
Order parameter harmonics,
$p_1$,
computed as a function of time parameter,
$\tau=D_{\ind{rot}} t$, by  numerically  solving
the initial value problem for 
the system~\eqref{eq:sys_p_n}
with the initial angular distribution~\eqref{eq:initial-distr}
at $\alpha_{\ind{ini}}=90$~deg
and $\alpha_{\ind{ini}}=0.0$~deg
for $-p_{\ind{ini}}=p_{u}\pm 0.05$. 
The photoexcitation and interaction parameters are:
$v_1=1.3$ and $v_2=-5$.
The stationary states are characterized by
the steady state harmonics:
$p_{\st}^{(-)}\approx -0.91$,
$p_{\st}^{(+)}\approx 0.77$
and $p_{u}\approx 0.48$. 
}
\label{fig:p1-90}
\end{figure}
 
In our modeling,
we numerically solve the initial value problem
for the harmonics $\tilde{\rho}_k$
with the initial angular distribution
\begin{align}
  \label{eq:initial-distr}
  \tilde{f}_{\ind{ini}}\propto \exp[-v_{\ind{ini}}\cos2(\phi-\phi_{\ind{ini}})].
\end{align}
The distribution~\eqref{eq:initial-distr} is characterized
by the initial value of
the order parameter harmonics
$\rho_1^{(\ind{ini})}=p_{\ind{ini}}=-I_1(v_{\ind{ini}})/I_0(v_{\ind{ini}})$
that gives the coupling coefficient $v_{\ind{ini}}$.
At the initial instant of time, the harmonics are thus given by
\begin{align}
  \label{eq:ini-harmonics}
\tilde{\rho}_k(0)  =\exp[2ik\alpha_{\ind{ini}}]
(-1)^k I_{k}(v_{\ind{ini}})/I_{0}(v_{\ind{ini}}).
\end{align}
In particular, for the order parameter harmonics, 
we have
\begin{align}
  \label{eq:ini-order-hrmn}
\tilde{p}_1(0)  =
p_{\ind{ini}} \cos(2\alpha_{\ind{ini}}),
\quad
\tilde{q}_1(0)  =
p_{\ind{ini}} \sin(2\alpha_{\ind{ini}}).
\end{align}

The initial data defined in
Eqs.~\eqref{eq:initial-distr} and~\eqref{eq:ini-harmonics}
represent the angular distribution of azo-dye molecules
formed in the film after the preparatory stage of initial irradiation with
linearly polarized UV light, $\vc{E}_{\ind{ini}}$.
Note that,
when the polarization vector  $\vc{E}_{\ind{ini}}$
is parallel to the $y$ axis, 
the angle $\alpha_{\ind{ini}}$ is equal to
the polarization azimuth of the reorienting light 
$\alpha_p$.

The system~\eqref{eq:sys_p_n} 
with the coupling coefficient $v=\tilde{v}=v_1+v_2 \tilde{\rho}_1$
describes how the harmonics evolve in time
approaching, in the long time limit,  
the steady state values
given in Eq.~\eqref{eq:pk_st_azim}.
These values are determined by the steady state order
parameter harmonics that can be found by solving 
the self-consistency equation~\eqref{eq:self_cosist_azim}

At sufficiently large positive photoexcitation parameter, 
$v_1>|v_c|$, the equilibrium (photosteady) state is unique and is characterized
by  the steady state order parameter harmonics $p_{\st}^{(-)}$.
This case is illustrated in figure~\ref{fig:rho1-65-mono}
that shows how the order parameter harmonics,
$\tilde{\rho}_1=p_1+i q_1$,
relaxes to the equilibrium state,
$p_1(t)\to p_{\st}^{(-)}<0$ and $q_1(t)\to 0$ at $t\to \infty$.
The equilibrium value of 
the principle (easy) axis angle~\eqref{eq:alph_e} then equals
the polarization azimuth $\alpha_p$ and 
$\alpha_e(t)\to \alpha_p$.
The latter implies that, in the regime of photosaturation,
the azo-dye molecules align perpendicular to the polarization vector
of the reorienting, 
$\uvc{n}_e\to (\cos\alpha_p,\sin\alpha_p,0)\perp\vc{E}_{UV}$.   
Such behavior is in complete agreement with
our experimental data described in Sec.~\ref{sec:experiment}
and has been previously observed in a number of 
experimental studies
(see, e.g., the monograph~\cite{Chigrin:bk:2008} and references therein).

Figure~\ref{fig:rho1-65-mono} additionally illustrates that
the magnitude of the order parameter harmonics, $|\rho_1|$,
related to the scalar order parameter, $s_{+}$,
given in Eq.~\eqref{eq:s_st} 
can be a nonmonotonic function of time.
For the angular distribution function
\begin{align}
 \label{eq:ang-distr-func}
  f(\phi,\tau)=(2\pi)^{-1} \sum_{k=-\infty}^{\infty}
  \tilde{\rho}_k(\tau) \exp[ -2 i k (\phi+\alpha_p)]
\end{align}
this implies that the distribution may experience broadening
at the early stage of relaxation
whereas it is getting narrower during the subsequent
relaxation to the highly ordered photosteady state.
The effect of early stage disordering can be seen in
Fig.~\ref{fig:distr-65-mono} that shows how
the orientational distribution function~\eqref{eq:ang-distr-func}
of azo-dye molecules evolves in time.
Note, however, that such transient disordering is typically 
either less pronounced or completely suppressed.
The latter case is demonstrated in 
Fig.~\ref{fig:distr-bstb}. 

Temporal evolution of the angular distribution
shown in Fig.~\ref{fig:distr-bstb} is computed
in the \textit{bistability region} where
the intermolecular interaction parameter, $v_2$,
is below its critical value, $v_2^{(c)}=-2$,
and the photoexcitation parameter, $v_1>0$,
is sufficiently small, $v_1<|v_c|$
(see Eq.~\eqref{eq:bimodal}). 
This is the region where,
in addition to the equilibrium state with
the negative order parameter harmonics $\tilde{\rho}_1=p_{\st}^{(-)}$,
the metastable state with the harmonics of opposite sign 
$\tilde{\rho}_1=p_{\st}^{(+)}$ is formed.

\begin{figure}[!htb]
\centering
\resizebox{120mm}{!}{\includegraphics*{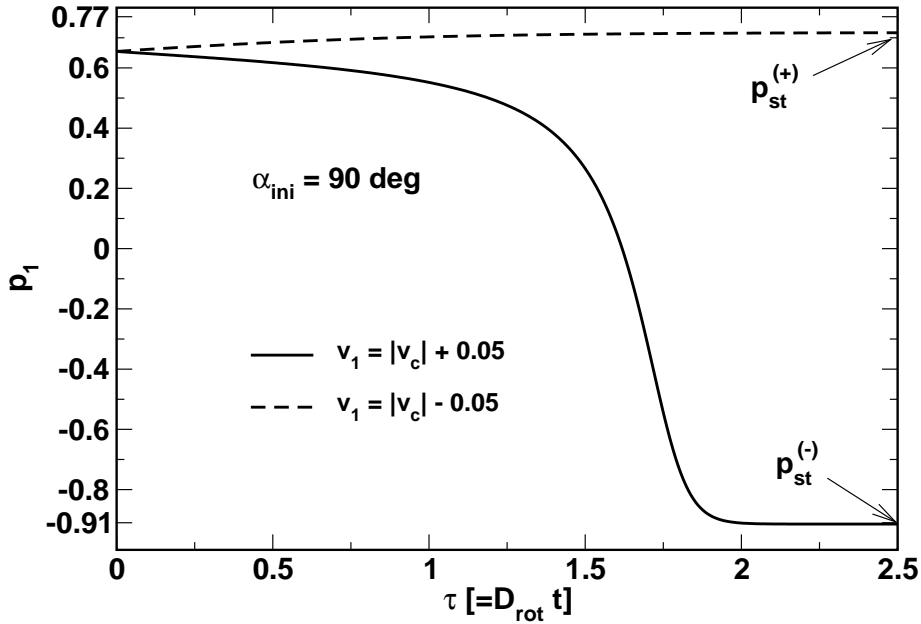}}
\caption{%
Order parameter harmonics,
$p_1$,
computed as a function of time parameter,
$\tau=D_{\ind{rot}} t$, by  numerically  solving
the initial value problem for 
the system~\eqref{eq:sys_p_n}
with the initial angular distribution~\eqref{eq:initial-distr}
at $\alpha_{\ind{ini}}=90$~deg
and $-p_{\ind{ini}}=0.65$
for $v_{1}=|v_{c}|\pm 0.05$. 
The interaction parameter 
and the critical value of the photoexcitation parameter are:
$v_2=-5$ and $|v_c|\approx 1.51$, respectively.
}
\label{fig:p1-thresh}
\end{figure}

Hence the stationary state free energy~\eqref{eq:F_st_azim}
takes the form of double-well potential
with two local minima located at 
$p_{\st}^{(-)}$ and $p_{\st}^{(+)}$.
There is also a local maximum
peaked at the point $p_{u}$ 
in the interval ranged from
$p_{\st}^{(-)}$ to $p_{\st}^{(+)}$.
Referring to Fig.~\ref{fig:bifur}(a),
at $v_1>0$,
this point with  $p_{u}>0$ lies on 
the upper half part of the branch
representing unstable stationary states.
In the case of real valued harmonics
with $q_k=0$,  
it is the boundary point 
of the basins of attraction of the two steady states.
In particular, this means that
the order parameter harmonics $p_1$
converges to either $p_{\st}^{(-)}$ or $p_{\st}^{(+)}$
depending on whether  its initial value $p_1(0)$
is below or above the boundary value $p_u$.

At
$\alpha_{\ind{ini}}=0$ and
$\alpha_{\ind{ini}}=\pi/2$,
the initial data~\eqref{eq:ini-harmonics}
are represented by two sets of 
real valued harmonics.
From Eq.~\eqref{eq:ini-harmonics}
it is evident that the only difference between 
the sets is the sign 
of the odd numbered harmonics .
In particular, we have $p_1(0)=p_{\ind{ini}}$ 
and $p_1(0)=-p_{\ind{ini}}$ 
at  
$\alpha_{\ind{ini}}=\alpha_p=0$ and
$\alpha_{\ind{ini}}=\pi/2$, respectively.

At $\alpha_{\ind{ini}}=0$ and $p_{\ind{ini}}<0$,
the harmonics characterizing the equilibrium state, 
$p_{\st}^{(-)}$,
is
the attracting point for the order parameter harmonics
and the azimuthal angle of the principal (easy) axis,
remains intact, $\alpha_e^{(\st)}=0$.
The latter comes as no surprise because
the polarization vectors of the initial irradiation
and of the reorienting light are parallel,
$\vc{E}_{UV}\parallel\vc{E}_{\ind{ini}}$ 

Interestingly, orientation of the easy axis
may stay unchanged even if the vectors, 
$\vc{E}_{UV}$ and $\vc{E}_{\ind{ini}}$,
are perpendicular
and $\alpha_{\ind{ini}}=\alpha_p=\pi/2$.
This occurs when the initial ordering is high,
so that $-p_{\ind{ini}}$ is greater than $p_u$,
$-p_{\ind{ini}}>p_u$.
As demonstrated in Fig.~\ref{fig:p1-90},
in this case,
the attracting point is 
the metastable harmonics $p_{\st}^{(+)}$
rather than the equilibrium one  $p_{\st}^{(-)}$
and $\alpha_e^{(\st)}=0$.

\begin{figure}[!htb]
\centering
\resizebox{120mm}{!}{\includegraphics*{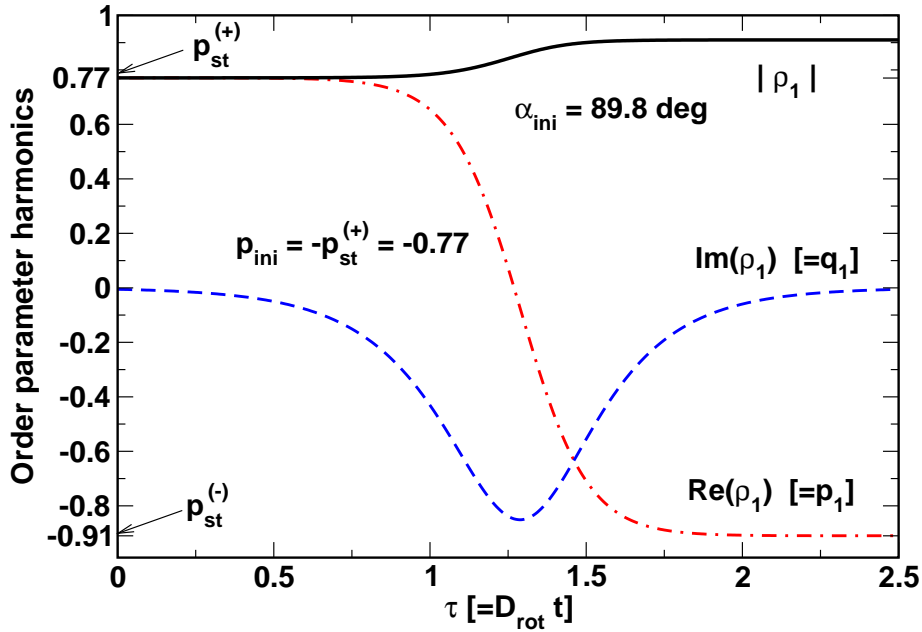}}
\caption{%
Order parameter harmonics,
$\rho_1$,
computed as a function of time parameter,
$\tau=D_{\ind{rot}} t$, 
at $\alpha_{\ind{ini}}=89.8$~deg
and $p_{\ind{ini}}=-p_{\st}^{(+)}$. 
Other parameters are listed in the caption of 
Fig.~\ref{fig:p1-90}.
}
\label{fig:rho1-89}
\end{figure}

It can also be seen from the curves of Fig.~\ref{fig:p1-90}
that, in the opposite case of
low ordered initial states with $-p_{\ind{ini}}<p_u$,
the order parameter harmonics $p_1$ relaxes back to
the equilibrium value $p_{\st}^{(-)}$.
So, at $-p_{\ind{ini}}<p_u$, the azimuthal angle, 
$\alpha_e^{(\st)}=\alpha_p=\pi/2$,
defines the steady state easy axis
which  is normal to the direction of its initial orientation.

In Fig.~\ref{fig:p1-thresh}, we show that,
when the initial ordering is sufficiently high,
the process of reorientation
crucially depends on whether
the photoexcitation parameter, $v_1$, 
which is proportional
to the intensity of the reorienting light, $I_{UV}$,
exceeds its critical value, $|v_c|$.
This critical value determines the intensity threshold,
$I_c= 4 |v_c/(3 u_I)|$, and
the easy axis reorientation is
suppressed at $\alpha_p=\pi/2$ 
in the low intensity region below the threshold where $I_{UV}<I_c$.

Similar threshold behavior
has been previously observed 
in a number of experimental studies
on light-induced easy axis reorientation
in nematic liquid crystal 
cells with photosensitive
polymer substrates~\cite{Dyad:jetpl:1992,Volosh:jjap:1995,Andrienko:mclc:1998,Chigrin:bk:2008}.
For azo-dye films, the  experimental data
of Refs.~\cite{Dubtsov:pre:2010,Dubtsov:apl:2012} 
also suggest the presence of a threshold
when the polarization vector $\vc{E}_{UV}$
is directed along the initial easy axis
(see the subsequent section for more details).
Our result that under certain conditions 
highly ordered initial distributions
may relax to the metastable stationary state
offers a possible explanation for this.

\begin{figure}[!htb]
\centering
\resizebox{120mm}{!}{\includegraphics*{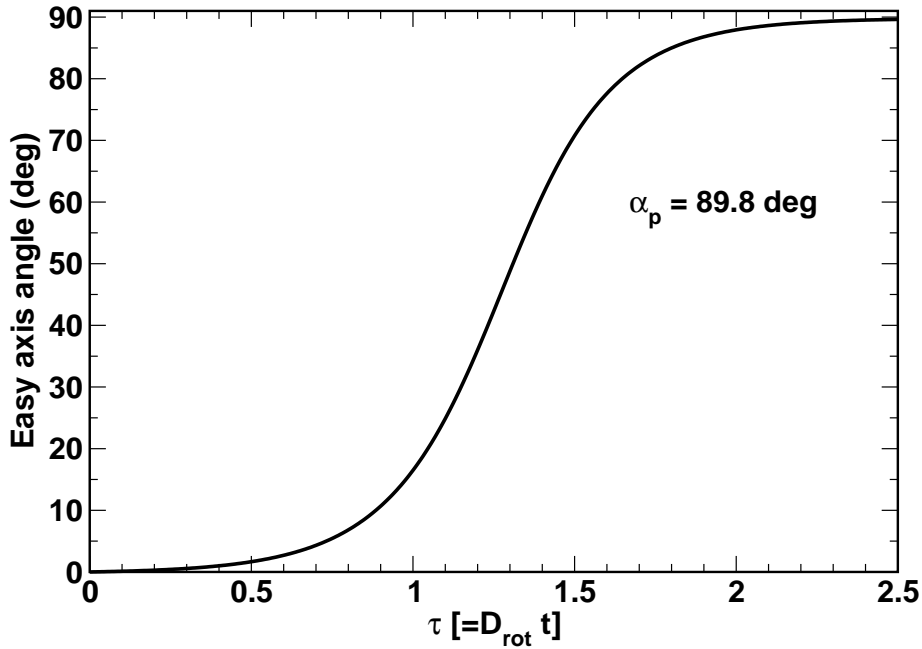}}
\caption{%
Principal axis azimuthal angle,
$\alpha_e$,
computed as a function of time parameter,
$\tau=D_{\ind{rot}} t$, 
at $\alpha_{\ind{ini}}=\alpha_{p}=89.8$~deg
and $p_{\ind{ini}}=-p_{\st}^{(+)}$.
Other parameters are listed in the caption of 
Fig.~\ref{fig:p1-90}. 
}
\label{fig:alpha-89}
\end{figure}

Referring to Fig.~\ref{fig:rho1-89}, 
it is seen that
small deviations of the angle $\alpha_{\ind{ini}}$
from $\pi/2$ have a destructive effect on
metastability of the attracting point
even if $-p_{\ind{ini}}=p_{\st}^{(+)}$.
It turned out that the basin of attraction of the metastable state
is unstable with respect to fluctuations of
the polarization azimuth $\alpha_p$.
In the vicinity
of the critical angle $\alpha_{\ind{ini}}=\pi/2$,
the presence of the metastable state
manifests itself as appreciably retarded 
relaxation to the equilibrium state
(see Fig.~\ref{fig:rho1-89}).
Similarly,
the shape of the azimuthal angle~\eqref{eq:alph_e} vs time 
curve shown in Fig.~\ref{fig:alpha-89}
points to a time delay in the process of 
principal (easy) axis reorientation.

\section{Theory versus experiment}
\label{sec:experiment}

In the previous section
we have examined the experimentally important
predictions of the model. 
These predictions can, in principle, be tested and,
 in this section, we discuss some of our recent experimental results
on
the reorientational dynamics of 
the electrically assisted light-induced 
azimuthal gliding of the easy axis at
varying polarization 
azimuth of LPUV reorienting light~\cite{Dubtsov:apl:2012}.

Such gliding is found to take place
on photoaligned azo-dye layers
when irradiation of nematic LC (NLC)
cells with LPUV light is combined with the application 
of ac in-plane electric
field~\cite{Pasechnik:lc:2008,Dubtsov:pre:2010}.
Since the effects of electric field is well beyond the scope 
of the theoretical approach under consideration, 
the regime of purely light-induced reorientation
will be our primary concern and
we deal with the data measured in the electric field-free regions

In our experiments, LC cells ($d=17.4 \pm 0.2\,\mu$m) 
of sandwich like type were assembled 
between two amorphous glass plates. 
The upper glass plate was
covered with a rubbed polyimide film 
giving the strong planar anchoring conditions.
 A film of the azobenzene sulfric dye SD1 (Dainippon Ink and
Chemicals)~\cite{Chigrin:bk:2008} 
was deposited onto the bottom substrate
on which transparent indium tin oxide (ITO) electrodes were placed.
The electrodes and the interelectrode stripes
(the gap was about  $g=50~\mu$m and
the in-plane ac voltage was $U=100$~V with $f=3$~kHz)
were arranged to be
parallel to the direction of  rubbing (the $x$ axis). 

The azo-dye SD1 layer was initially illuminated by LPUV
light at the wavelength $\lambda = 365$~nm.
The preliminary irradiation produced the zones of different energy dose exposure 
$D_p=0.27,\,0.55$~J/cm$^2$ characterized by relatively 
weak azimuthal anchoring strength. 
The light propagating along 
the normal to the substrates (the $z$ axis) 
was selected by an interference filter.
Initial orientation of the polarization vector of the actinic light,
$\mathbf{E}_{\mathrm{ini}}$, was chosen so as to
align azo-dye molecules at a small angle of 4 degrees to 
the $x$ axis, $\varphi_0\approx 4$~deg. 

The LC cell was filled with the nematic LC mixture E7 (Merck) 
in isotropic phase and then slowly cooled down to 
room temperature. 
Thus we prepared the LC cell
with a weakly twisted planar orientational structure 
where the director at the bottom surface 
$\mathbf{n}_0$ is clockwise rotated  
through the initial twist angle $\varphi_0\approx 4$~deg
which is the angle between $\mathbf{n}_0$ 
and the director at the upper substrate (the $x$ axis).

In addition to the electric field,
$E=2$~V/$\mu$m,
the cell was irradiated with the reorienting
LPUV light beam 
($I_{UV}=0.26$~mW/cm$^2$ and $\lambda = 365$~nm)
normally impinging onto the bottom substrate.
For this secondary LPUV irradiation,
orientation of the polarization plane is determined by
the polarization azimuth, $\alpha_p$. 
 
Our experimental method has already been described 
in Refs.~\cite{Pasechnik:lc:2008,Dubtsov:pre:2010}.
In this method, NLC orientational structures 
were observed via a polarizing microscope connected
with a digital camera and a fiber optics spectrometer. 
The rotating polarizer technique was used to measure 
the azimuthal angle
$\varphi_e$ characterizing orientation of the easy axis. 
In order to register microscopic images and 
to measure the value of $\varphi_e$,
the electric field and the reorienting light were switched off 
for about 1 min.
This time interval is short enough 
to ensure that orientation of the easy axis remains essentially intact
in the course of measurements.
The measurements were carried out at a temperature of 26$^{\circ}$C.

\begin{figure}[!htb]
\centering
\resizebox{100mm}{!}{\includegraphics*{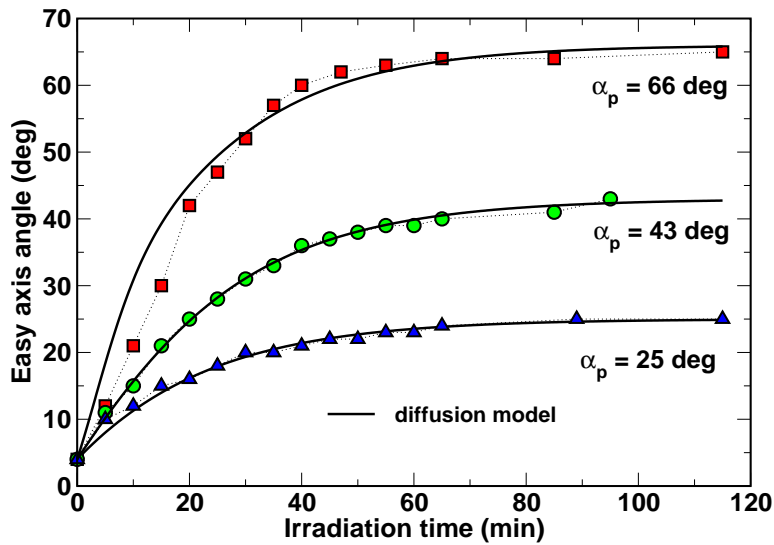}}
\caption{%
Easy axis angle versus the irradiation time
at different values of
the polarization azimuthal angle
of  the reorienting  UV light 
$\alpha_p$.
The experimental data
were measured by using 
the linearly polarized activating light
with $\lambda=365$~nm and $I=2.6$~W/m$^2$.
Solid lines represent the theoretical curves
computed by  numerically  solving
the initial value problem for 
the system~\eqref{eq:sys_p_n}
with the initial angular distribution~\eqref{eq:initial-distr}.
The parameters used in calculations are as follows:
the photoexcitation (interaction) parameter is
$v_1=1.3$ ($v_2=-5$); 
$p_{\ind{ini}}=p_{\st}^{(-)}$
($\alpha_p=25$~deg and $\alpha_p=43$~deg)
and
$p_{\ind{ini}}=-0.25$ at $\alpha_p=65$~deg.
Other parameters are listed in the caption of 
Fig.~\ref{fig:p1-90}.
}
\label{fig:fit-diff-bstb}
\end{figure}


The case
where the reorienting light is linearly polarized
along the initial surface director $\mathbf{n}_0$ and $\alpha_p=\pi/2$, 
was studied in Refs.~\cite{Pasechnik:lc:2008,Dubtsov:pre:2010}.
It turned out that, by contrast
to the electrically assisted light-induced gliding, 
the purely photoinduced reorientation
is almost entirely inhibited at $\alpha_p=\pi/2$.
As it can be seen from Fig.~\ref{fig:fit-diff-bstb},
the latter is no longer the case
for the reorienting light with $\alpha_p\ne \pi/2$.

Thus the result is that,
as opposed to the cases 
with non-vanishing electric field and $\alpha_{p}\ne\pi/2$,
the purely photoinduced reorientation of the easy axis 
has been completely suppressed when
the polarization plane is parallel to the initial easy axis.
From the above discussion,
our model predicts similar behavior 
that occurs in the bistability region where
the reorientation is characterized by
the threshold for the intensity of 
reorienting light.

Figure~\ref{fig:fit-diff-bstb}
shows the easy axis angle
measured  as a function of the irradiation time  
at various values of the polarization azimuth.
The measurements were performed
in the region outside the interelectrode gaps  
where the ac electric field is negligibly small.
In the zero-field curves,
the easy axis angle increases with
the irradiation time
starting from the angle of initial twist,
$\varphi_0$,
and
approaches the photosteady state
characterized by the photosaturated value of the
angle close to $\alpha_p$. 

The experimental data
can be fitted by the theoretical curves
computed from the formula~\eqref{eq:alph_e}, 
where the order parameter harmonics,
$\tilde{\rho}_1$, is found by  numerically  solving
the initial value problem for 
the system~\eqref{eq:sys_p_n}
with the initial angular distribution~\eqref{eq:initial-distr}.
In Fig.~\ref{fig:fit-diff-bstb}
we make
a comparison between the experimental data and the theoretical curves 
computed in the bistability region
by setting 
the photoexcitation and interaction parameters, $v_1$ and $v_2$, 
equal to $1.3$ and $-5$, respectively.
In this region, the equilibrium state 
is characterized by the steady state harmonics
$p_{\st}=p_{\st}^{(-)}\approx -0.91$,
whereas the stationary harmonics $p_{\st}^{(+)}\approx 0.77$
corresponds to the metastable state.
For $\alpha_p=25$~deg and $\alpha_p=43$~deg,
the curves were computed at $p_{\ind{ini}}=p_{\st}^{(-)}$. 
It follows that $|\rho_1(0)|=|\rho_{\st}|$
and the angle $\alpha_{\ind{ini}}$ 
is assumed to be the only parameter 
describing difference between the initial and the equilibrium 
angular distributions. 
The rotational diffusion coefficient
$D_{\ind{rot}}$ can be estimated at about 
$1.6\times 10^{-4}$~s$^{-1}$
Using this value to fit the data measured at $\alpha_p=65$~deg,
we find that the theoretical and experimental results are in
agreement when the initial ordering parameter
$p_{\ind{ini}}\approx -0.25$ differs 
from the equilibrium value.

\begin{figure}[!htb]
\centering
\resizebox{100mm}{!}{\includegraphics*{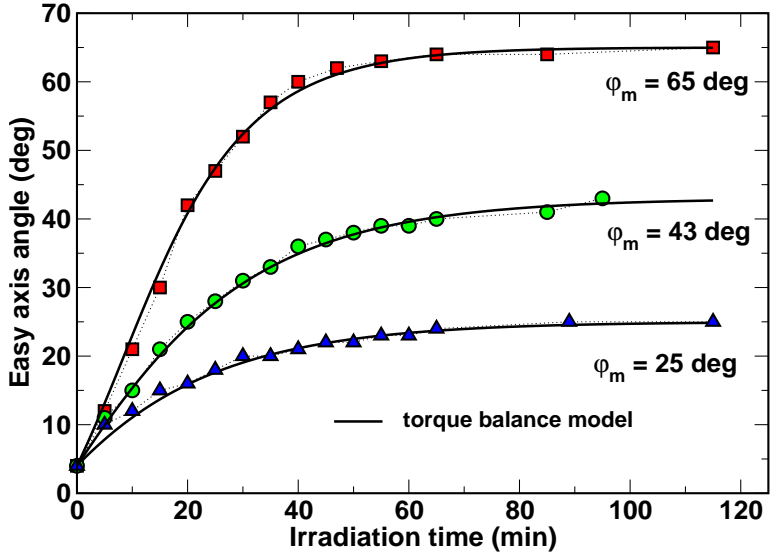}}
\caption{%
Easy axis angle versus the irradiation time
in the electric field-free region
at different values of
the polarization azimuthal angle
of  the reorienting  UV light.
The experimental data
were measured by using 
the linearly polarized activating light
with $\lambda=365$~nm and $I=2.6$~W/m$^2$.
Solid lines represent the theoretical curves
computed from the formula~\eqref{eq:phie-sol}
at $t_e=\gamma_e/W_e=23$~min 
(for $\varphi_m=25$~deg and $\varphi_m=43$~deg),
and $t_e\approx 14.4$~min.
for $\varphi_m=65$~deg.
}
\label{fig:fit-torq-0}
\end{figure}

We conclude this section
with the remark
concerning the phenomenological model
formulated in Refs.~\cite{Pasechnik:lc:2006,Pasechnik:lc:2008}
to describe the effect of the electrically assisted photoinduced
gliding of the easy axis.
According to this model,
in the absence of ac electric field,
temporal evolution of the easy axis azimuthal angle
$\varphi_e$ can be described by the balance torque
equation~\cite{Dubtsov:apl:2012}
\begin{align}
  \label{eq:torque-phie}
  \gamma_e
\pdr{\varphi_e}{t}=-\frac{W_e}{2}
\sin2 (\varphi_e-\varphi_m),
\end{align}
where 
$\gamma_e$ is \textit{the specific viscosity of gliding};
$W_e$ is 
the effective anchoring parameter which defines 
the strength of coupling between
the easy axis $\vc{n}_e$ and
the initial state of surface orientation;
$\varphi_m$ is the polarization dependent
phase shift.
The solution of the dynamic equation~\eqref{eq:torque-phie}
\begin{align}
  \label{eq:phie-sol}
  \tan(\varphi_e(t)-\varphi_m)=\tan(\varphi_e(0)-\varphi_m) \exp[-t/t_e],
\end{align}
where $t_e=\gamma_e/W_e$
the characteristic time  of 
easy axis reorientation,
can now be used to fit the experimental data.
The results are presented in Fig.~\ref{fig:fit-torq-0}.

From the curves 
depicted in Fig.~\ref{fig:fit-torq-0}
it can be inferred that 
the phase shift $\varphi_m$
is close to the polarization azimuth
of the reorienting light.
The value of the characteristic time,
$t_e\approx 14.4$~min,
estimated for $\varphi_m=65$~deg
differs from the one $t_e\approx 23$~min
obtained for other two angles,
$\varphi_m=25$~deg and $\varphi_m=43$~deg.
From the other hand,
in Fig.~\ref{fig:fit-diff-bstb},
the case with $\alpha_p\approx 65$~deg
is distinctive in initial ordering of the film. 
This result points to the fact that 
the coupling strength of the easy axis $W_e$
depends on the photoinduced order parameter
of  the initially irradiated azo-dye layer.
Such dependence was 
one of the key assumptions taken
in Ref.~\cite{Dubtsov:pre:2010}
when applying
the phenomenological model~\cite{Pasechnik:lc:2006,Pasechnik:lc:2008}
to interpret the experimental data
for photoaligned azo-dye layers 
prepared at different initial irradiation doses.


\section{Discussion and conclusions}
\label{sec:conclusion}
 
In this paper we have studied the kinetics of
orientational reordering 
that takes place in the azo-dye films
under the action of LPUV reorienting light.
In our theoretical approach,
the process of reordering
is described
in terms of a rotational Brownian motion
and  we have employed the 2D diffusion model
formulated in Ref.~\cite{Kis:pre:2009}
to explore the peculiarities of the photoinduced reordering in
the regime of purely in-plane reorientation.

Orientational ordering of azo-dye molecules
is characterized by the in-plane (2D) 
order parameter tensor~\eqref{eq:S_2-def}.
This tensor is
expressed in terms of the order parameter
harmonics, $\rho_1$, 
that enters the formula~\eqref{eq:S_2-eigval}
giving both the value of 
the scalar order parameter, $s_{+}$,
and the azimuthal angle, $\alpha_e$, 
of the principal (easy) axis, $\uvc{n}_e$.

It is found that dynamics of the order parameter 
harmonics essentially depends on 
the initial ordering and the polarization azimuth
of the LP reorienting light.
In the long time limit, 
the order parameter harmonics approaches
its steady state value reaching the regime of 
photosaturation.
We have proved 
that, in the steady state,
the easy axis azimuthal angle is 
dictated solely by the polarization azimuth, $\alpha_p$,
(see Eq.~\eqref{eq:s_alph_st}),
whereas the photostationary value of 
the scalar order parameter is determined
by the photoexcitation and intermolecular 
interaction parameters, $v_1\propto I_{UV}$
and $v_2$, defined in Eq.~\eqref{eq:v1_v2}.

In the $v_1$-$v_2$ plane,
inequalities~\eqref{eq:bimodal}
define the bistability region
(see Fig.~\ref{fig:bifur}(b))
with two locally stable stationary states:
equilibrium and metastable ones.
An experimentally important bistability effect
is that the easy axis reorientation
is inhibited when 
the intensity of the reorienting light linearly 
polarization along the initial easy axis 
is below its critical (threshold) value.
Such threshold behavior
was observed in our experiments
where the easy axis angle was measured
as a function of the irradiation time
at varying polarization azimuth of LPUV reorienting light.

An alternative phenomenological model that
also predicts the presence of the reorientation threshold
was suggested in Ref.~\cite{Alexe:pre:2007}.
In that model, the threshold effect 
comes about from
the interplay between two competing easy axes.
By contrast, in our model, 
the bistability appears as  an intrinsic property
of the reordering process in azo-dye films
that manifests itself as the threshold of reorientation.  

There is another interesting bistability effect
that occurs
when the light is switched off ($I_{UV}=0$) and $v_2<v_2^{(c)}=-2$.
In this case,
owing to the rotational symmetry of the two-particle 
interaction~\eqref{eq:U_2},
the easy axis angle, $\alpha_e$, becomes undetermined
and plays the role of the degeneracy parameter.
The LPUV light field lifts the degeneracy and 
the dynamics of the angle turned out to be predominately governed
by the photoexcitation parameter, $v_1$.

For the scalar order parameter
which is related to the magnitude of the order parameter harmonics, $|\rho_1|$,
the rate of relaxation is typically at least twice faster than that 
for the easy axis angle.
As is illustrated in Fig.~\ref{fig:rho1-65-mono},
depending on the initial conditions,
temporal evolution of this order parameter can be non-monotonic.
These effects can also be seen from
Figs.~\ref{fig:distr-65-mono} and~\ref{fig:distr-bstb}
that show how the orientational angular distribution of azo-dye molecules
evolves in time.
 
By
numerically solving the dynamical system
for the angular distribution harmonics~\eqref{eq:sys_p_n}
we have fitted the experimental data
on the light-induced azimuthal gliding of
the easy axis at the photoaligned azo-dye SD1 layer.
In our analysis,
the photoinduced reordering of azo-dye molecules
was assumed to be the key factor
dominating the easy axis reorientation. 

Despite the fact that good agreement between the computed curves
and the data (see Fig.~\ref{fig:fit-diff-bstb}) 
counts in favor of our simplified approach,
a deeper insight into complicated
molecular mechanisms behind the gliding
under consideration 
requires additional experimental studies and a 
more systematic theoretical treatment 
of the processes involved.
Such treatment
has to deal with the interplay of photoinduced ordering
in azo-dye films,
the anchoring energy effects~\cite{Kis:pre2:2005}
and the adsorption-desorption 
processes~\cite{Vetter:jjap:1993,Ouskova:pre:2001,Barbero:bk:2006,Fedorenko:pre:2008}
underlying the gliding phenomenon.

This work was partially supported by grants: Development of the Higher
School’s Scientific Potential 2.1.1/5873; Grant NK-410P; HKUST CERG
RPC07/08.EG01 and CERG 612208, 612409.


%

\end{document}